\documentclass{emulateapj}

\newcommand{\beq}	{\begin{equation}}
\newcommand{\eeq}	{\end{equation}}
\newcommand{\beqa}	{\begin{eqnarray}}
\newcommand{\eeqa}	{\end{eqnarray}}
\newcommand{\calm}	{{\cal M}}
\newcommand{\mrms}      {{\calm_{\rm rms}}}

\newcommand{\vecB}	{{\bf B}}

\newcommand{\veck}	{{\bf k}}
\newcommand{\vecv}	{{\bf v}}

\newcommand{\vecnabla}	{{\bf\nabla}}
\newcommand{\grad}      {{\bf\nabla}}
\newcommand{\divb}	{{\grad \cdot \vecB}}

\newcommand{\alfven}    {{Alfv$\acute{\rm e}$n }}

\newcommand{\vrms}	{v_{\rm rms}}

\newcommand{\avg}[1]    {{\langle #1 \rangle}}

\newcommand{\ma}	{{\calm_{\rm A}}}
\newcommand{\mao}	{{\calm_{\rm A,0}}}

\newcommand{\tff}	{t_{\rm ff}}

\newcommand{\kmax}	{k_{\rm max}}
\newcommand{\ravg}	{\avg{R}}

\shorttitle{MHD Turbulence Resolution and Refinement Study with AMR}
\shortauthors{Li, McKee, \& Klein}
\begin{document}

\title{A Stable, Accurate Methodology for High Mach Number, Strong Magnetic Field MHD Turbulence with Adaptive Mesh Refinement: Resolution and Refinement
Studies}
\author{Pak Shing Li}
\affil{Astronomy Department, University of California,
    Berkeley, CA 94720}
\email{psli@astron.berkeley.edu}
\author{Daniel F. Martin}
\affil{Lawrence Berkeley National Laboratory, 1 Cyclotron Road, Berkeley, CA 94720}
\email{DFMartin@lbl.gov}
\author{Richard I. Klein}
\affil{Astronomy Department, University of California, Berkeley, CA 94720; \\
    and Lawrence Livermore National Laboratory, P.O.Box 808, L-23, Livermore, CA 94550}
\email{klein@astron.berkeley.edu}
\and
\author{Christopher F. McKee}
\affil{Physics Department and Astronomy Department, University of California,
    Berkeley, \\
    CA 94720}
\email{cmckee@astro.berkeley.edu}

\begin{abstract}
Performing a stable, long duration simulation of driven MHD turbulence with a high thermal Mach number and a strong initial magnetic field is a challenge to high-order Godunov ideal MHD schemes because of the difficulty in guaranteeing positivity of the density and pressure.  We have implemented a robust combination of reconstruction schemes, Riemann solvers, limiters, and Constrained Transport EMF averaging schemes that can meet this challenge, and using this strategy, we have developed a new Adaptive Mesh Refinement (AMR) MHD module of the ORION2 code.
We investigate the effects of AMR on several statistical properties of a turbulent ideal MHD system with a thermal Mach number of 10 and a plasma $\beta_0$ of 0.1 as initial conditions; our code is shown to be stable for simulations with higher Mach numbers ($\mrms = 17.3$) and smaller plasma beta ($\beta_0 = 0.0067$) as well.
Our results show that the quality of the turbulence simulation is generally related to the volume-averaged refinement.  Our AMR simulations show that the turbulent dissipation coefficient for supersonic MHD turbulence is about 0.5, in agreement with unigrid simulations.
\end{abstract}

\section{Introduction}
Eulerian codes are commonly used in star formation studies in order to
model the complex physical processes involved, including turbulence,
magnetic fields, gravitational collapse, and radiation feedback.  The
dynamic ranges of density and size scales involved in star formation are
enormous, ranging from more than 10 pc in giant molecular clouds (GMCs)
of density $\sim 5$~AU protostellar cores with densities $\ga 10^{13}$ 
cm$^{-3}$ \citep{mas98}.  This poses a significant challenge to
numerical simulations using a uniform computational mesh.  For example, using
the unigrid code ZEUS-MP \citep{hay06}, a simulation of ideal MHD turbulence with
a $1024^3$ grid requires $\sim 50,000$ cpu hours.  When gravitational collapse begins, dense cores will reach the numerical resolution limit \citep{tru97} in just a small fraction of the global free-fall time, $\tff$.  For a high-order Godunov scheme, the computing time will be $\sim$~5 times that for ZEUS-MP, which uses a low-order scheme.  The computing time is further increased by a factor of at least 16 whenever the resolution of the 3D grid is doubled because maximum \alfven speed will be increased at lower density regions as the result of higher resolution.  It is computationally inefficient to simply increase the grid resolution for star formation simulations  because
only a small fraction of the simulated region has collapsed to sufficiently
high density to violate  numerical resolution requirements; most of the simulated volume is in low density voids where such fine resolution is unnecessary. Therefore, adaptive mesh refinement (AMR) becomes an important tool for simulation of star formation using Eulerian codes.  With AMR, the computational mesh is refined only in the localized regions where high resolution is required, and as a result computational resources are concentrated in the regions where they are needed most.

Stars form in molecular clouds, which are cold ($T\sim 10$~K), supersonically  turbulent (sonic Mach numbers $\sim 10$ on scales $\sim 10$ pc), and magnetized ($B\ga 10 \mu$G); the \alfven Mach number is observed to be of order unity and the plasma $\beta$ parameter is small ($\la0.1$; \citealp{cru99}).  There are several MHD codes with AMR capability, including the publicly available codes Ramses \citep{tey02}, PLUTO \citep{mig07}, ENZO \citep{wan09}, and FLASH \citep{fry00}.  However, to our knowledge, there is no AMR code in the literature that has demonstrated the capability of simulating supersonic MHD turbulence with initial conditions appropriate for star-forming regions.  The primary reason for this is that Godunov schemes for ideal MHD with high-order approximate Riemann solvers cannot guarantee positivity in density and pressure \citep[e.g.][]{lev92,tor99,ber05,zha10} and are therefore unstable for turbulence that is driven for long times with such initial conditions.  This becomes an important consideration when developing an AMR ideal MHD code for star formation simulations with MHD turbulence.

Because turbulence is intermittent, one might hope that AMR would be very effective in simulating it.  However, in a study of purely hydrodynamic turbulence using the ENZO code, \citet{kri06} argued that the use of AMR only became practical from an efficiency standpoint if the base mesh had high resolution to start with.  When they attempted runs with coarse ($128^3$ and $256^3$) base meshes (note that they used refinement ratios of 4), a large fraction of their domain was refined until they were sufficiently able to resolve the localized turbulent structures using AMR: with the refinement criterion they adopted, they refined 90\% of the computational domain at $512^3$ resolution, 65\% of the domain at $1024^3$ resolution and 34\% at $2048^3$ resolution.
Furthermore, it is not just a matter of whether AMR is economical for a turbulence simulation, but also whether it can accurately capture the properties of the turbulence.  Another attempt at using different refinement criteria, such as refining on local vorticity and divergence of velocity in addition to shock refinement for purely hydrodynamic turbulence, also shows that very large refinement coverage generally results in capturing the turbulence statistics \citep{sch09}.

In this paper, we present a robust MHD AMR scheme that is able to simulate turbulent flows with high Mach numbers and strong initial magnetic fields and that uses an accurate CT scheme to maintain $\vecnabla\cdot\vecB=0$ to machine accuracy without recourse to approximate methods that rely on either divergence cleaning\citep[e.g.][]{cro05} or monopole advection \citep[Powell's 8-wave scheme and 
its extensions; e.g.][]{pow99,ded02,mig07,wan09,waa11}.  Recently, \citet{waa09} modified the MHD module of the FLASH code using a directionally split MUSCL-Hancock scheme with properly discretized Powell source terms in order to enable stable driven-turbulence simulations.  The  driven-turbulence tests in \citet{waa11} are either at high Mach numbers with a relatively weak initial magnetic field ($\beta_0 \sim 1$) or at low thermal Mach numbers ($\la 2$) with somewhat stronger fields ($\beta_0 \sim 0.25$).  The tests they carried out were all unigrid; the performance of their code on driven turbulence with AMR was not described.  \citet{waa11} cites simulations of driven turbulence at high Mach numbers in strong fields using this code, but these too were unigrid simulations.

Here we present the results of a quantitative investigation of simulations of ideal MHD turbulence using a newly implemented ideal MHD module in our AMR code, ORION2, which is based on a conservative high-order Godunov scheme. We investigate the significance of refinement coverage to the quality of turbulence statistics in strongly supersonic MHD turbulent flow.  We are able to perform long duration, high Mach number, driven MHD turbulence simulations with a magnetic field that is initially moderately strong ($\beta_0 = 0.1$).  In \S 2, we briefly describe the numerical method and the implementation of our AMR constrained transport scheme using the Chombo AMR framework \citep{col00}.  In \S 3, we present several standard tests to examine the accuracy of the code for both unigrid and AMR simulations.  In \S 4 we discuss the effects of refinement coverage on several ideal MHD turbulence statistics using AMR.  We focus our discussion on the velocity power spectrum, the PDF, and the turbulence dissipation rates.  In \S 5 we present our conclusions.

\section{Numerical Method}
\subsection{ORION2}

In this paper we present ORION2, which represents a major upgrade of our parallel 
radiation hydrodynamic AMR code "ORION" \citep{tru98,kle99,kru04,kru07}.
ORION2 is implemented using the Chombo AMR framework \citep{col00} and is in a modular form that allows the selection of a variety of physics modules for simulations.  Chombo is a set of highly optimized tools that provide an infrastructure for implementing finite difference and finite volume methods for solving partial differential equations on a block-structured AMR grid configuration.  Elliptic and time-dependent modules are included, as well as support for parallel platforms using MPI.  The newly developed MHD module of ORION2 is based on the Godunov scheme in the finite volume formalism implemented in the PLUTO code \citep{mig07}.  Specifically, we use a dimensionally unsplit Corner Transport Upwind (CTU) scheme \citep{col90} incorporating the Constrained Transport (CT) framework \citep{eva88}.  This CTU+CT integrator is best described in \citet{sto08}, which will not be repeated here.   Our implementation preserves some of the flexibility of the PLUTO code in choosing (1) different interpolation schemes to re-construct cell interface states from the cell-center values, such as the piecewise linear method (PLM) or the piecewise parabolic method (PPM); (2) different limiters for the preservation of monotonicity near a discontinuity during the re-construction stage, from the very diffusive minmod to least diffusive monotonized central difference limiter; (3) different Riemann solvers, such as Roe and HLL-family solvers, to obtain fluxes at the cell interfaces based on the reconstructed cell interface states; and (4) different CT electromotive force (EMF) averaging schemes, such as the simple arithmetic averaging of fluxes computed during the upwind step \citep{bal99}, or a face-to-edge integration procedure using the arithmetic average of the EMF derivatives from neighbor cells,
or selecting EMF derivatives according to the sign of the mass flux at the cell interface.  See \citet{gar05} on how to compute the EMF at cell edges during the upwind step.  \citet{mig07} have summarized the advantages and disadvantages of the combinations of different solvers and integration schemes.

The CT scheme increases the complexity of the algorithm, especially for an AMR
code, and it requires additional memory for storing the face-centered magnetic
field. However, the benefit of the CT scheme is that the solenoidal constraint
$\divb = 0$ is ensured to machine accuracy. For cell-centered MHD algorithms,
divergence-cleaning methods \citep[e.g.][]{cro05} are used to ensure the solenoidal constraint, but they cannot guarantee positivity of pressure (or energy) and therefore reduce the robustness of the code.  The \citet{ded02} approach, which is used in some cell-centered codes \citep[e.g.][]{pen09,mig10},
allows magnetic monopoles to decay with time as they are transported to the domain boundaries.  Some studies have reported that monopoles could lead to incorrect jump conditions and other spurious dynamical effects \cite[e.g][]{bal99,tot00}.  The cell-centered field is readily calculated with the CT scheme: once the face-centered magnetic field is updated, the cell-centered magnetic field is computed from a simple averaging of the face-centered magnetic fields.

\subsection{AMR Implementation}

Extending any uniform-mesh algorithm to AMR requires several modifications, primarily involving the coupling between the solutions at different
resolutions. We follow the block-structured AMR approach outlined by
\cite{bc89} and extended to MHD by \cite{bal01}. 

We begin with a uniform mesh that spans the domain and is denoted as AMR
level $\ell=0$, with mesh spacing $h^{0}$.  (Superscripts on the symbols $h$, $n_{\rm ref}$, $t$ and $\Delta t$ indicate the level of refinement, not a power.)
When refinement is triggered by some criterion, such as a steep density, velocity or magnetic field gradient, refined grids are constructed in logically-rectangular patches with mesh spacing $h^{1} = h^{0}/n_{\rm ref}^{0}$, which are grouped into AMR level 1; here the refinement ratio $n_{\rm ref}^{0}$ is some power of 2. If further refinement is desired, more AMR levels are constructed. Each AMR level $\ell$ has a uniform mesh spacing $h^\ell = h^{\ell-1}/n_{\rm ref}^{\ell-1}$.  In general, $n_{\rm ref}^\ell$ can have a non-uniform dependence on $\ell$, but in this paper we adopt $n_{\rm ref}^{\ell-1}=2$.  Patches with the same resolution are organized into levels, which are then organized into a hierarchy of AMR levels.  

Following the approach outlined by \cite{bc89}, we refine in time as well as space, commonly known as ``subcycling'' -- the solution on each AMR level is updated using a timestep $\Delta t^\ell = \Delta t ^{\ell-1}/n_{\rm ref}^{\ell-1}$. If we know the solution on the entire AMR hierarchy at time $t^0$, then we begin the multilevel update of the solution on all levels by updating the base level solution by $\Delta t^0$, without regard for any finer levels. We then recursively advance any refined levels until the solution on the entire AMR hierarchy has reached $t^0 + \Delta t^0$. Whenever the solutions on different levels reach the same solution time, they are ``synchronized'', to ensure that the composite solution remains conservative and maintains the solenoidal nature of the magnetic field. 

In general, the update of the solution on each logically-rectangular patch
follows the same approach as that for a uniform-mesh implementation. AMR-specific implementation details fall into the following categories:
\begin{itemize}
\item [1)]
{\bf Boundary conditions}:
Following common practice, boundary conditions at the edge of a rectangular
patch are handled by adding a ring of ghost cells around the patch sufficient
to complete the stencils used to update the solution on the interior of the
patch (otherwise known as the ``valid region''). The algorithm in this
work requires 4 ghost cells. Solution values in ghost cells (and the magnetic fields on the associated ``ghost faces'') are filled depending on the type of boundary they are associated with:
\begin{itemize}
\item [1.1)]
  {\it Physical domain boundaries} -- if the patch boundary abuts a
  (non-periodic) physical domain boundary, ghost cell values are set using
  standard discretizations of the relevant physical domain boundary condition
  (e.g. Dirichlet or Neumann boundary conditions). 
\item [1.2)]
  {\it Copy boundaries} -- if the patch is adjacent to other patches on
  the same level, ghost-cell values are filled by copying interior
  (valid-region) values from adjacent patches. 
\item [1.3)]
  {\it Coarse-fine boundaries} -- if the ghost cells are along a
  coarse-fine boundary between two AMR levels, coarse-level solution values
  are interpolated to fill in ghost regions. If the fine level is not at the
  same solution time $t$ as the coarse-level solution, we interpolate the
  coarse-level data linearly in time, since $t^{\ell-1}_{old} \leq t^\ell \leq
  t^{\ell-1}_{new}$. 
  Cell-centered conserved quantities
  are filled using piecewise linear interpolation that is limited to prevent
  the creation of new maxima and minima, following \cite{bc89}. The
  face-centered magnetic field is interpolated by first performing a
  piecewise-linear interpolation of the coarse-level field onto the fine-level
  faces that overlie the coarse-level faces, and then linearly interpolating
  these interpolated values in the face-normal direction to fill the fine-level
  faces that do not overlie coarse-level faces. Note that this differs from
  the divergence-free interpolation scheme presented by \cite{bal01} in that
  interpolated values are not necessarily divergence free.
   We have not found
  it necessary to ensure that the magnetic fields in ghost regions are
  divergence-free, because the interpolated ghost values are used only in the
  reconstruction scheme to compute the fluxes in the divergence-preserving CT
  scheme and are never directly used to increment the magnetic fields
  themselves. 
\end{itemize}

\item [2)]
{\bf Interpolation to newly-refined regions}:
As the solution evolves, the refined regions evolve with it through a regridding process.  Previously refined regions are de-refined when finer resolution is no longer needed, while coarse-level regions are refined when finer resolution is required.  In the de-refinement case, we simply average the fine-level solution to the newly-exposed coarse-level mesh. Newly-refined regions are filled by interpolating the coarse-level solution. Cell-centered conserved quantities are interpolated using piecewise linear interpolation that is limited to prevent the creation of new solution maxima and minima. The face-centered magnetic field is defined using a two-step process. First, we interpolate the coarse-level $\vecB$ field onto the new faces using the face-centered interpolation used to fill ghost faces at coarse-fine interfaces. Then, the newly interpolated values are projected
using a variant of the face-centered projection described in \cite{mar00} to ensure that they are discretely divergence-free.

\item [3)]
{\bf Synchronization}:
When the solutions on two AMR levels reach the same time, a series of
synchronization operations is performed. First, the fine-level solution is
averaged onto the covered regions of the coarser level. This includes both
cell-centered conserved variables and the face-centered $\vecB$. The coarse-
and fine-level solutions have been updated using fluxes that have been computed independently, likely resulting in a loss of conservation at coarse-fine interfaces.  Conservation is maintained through a flux-correction step similar to that used by \cite{bc89}, which ensures that the same fluxes are used to update the coarse- and fine-level solutions across coarse-fine interfaces. 

Similarly, the magnetic field is unlikely to be divergence-free at coarse-fine
interfaces because coarse- and fine-level magnetic fields have been updated
independently. Following \cite{bal01}, we ensure a divergence-free $\vecB$
at coarse-fine interfaces through a correction step that ensures that the
same electric field values are used to update the coarse- and fine-level
$\vecB$ field adjacent to coarse-fine interfaces.

\end{itemize}

\section{Standard Test Results}
\label{tests}
We have performed many tests of the ORION2 MHD module, including the well-known standard tests, such as the EM wave families \citep[e.g.][]{cro05}, the \citet{ryu95} shock-tube tests, the \citet{bri88} shock-tube test, the field-loop advection test, and the MHD blast-wave test as in \citet{gar05}.  We present the results of only two of the shock-tube tests and the field-loop advection test here to demonstrate the second-order accuracy of the MHD module.

\subsection{Shock-tube Tests}
\begin{figure}
\includegraphics[scale=0.7,angle=-90]{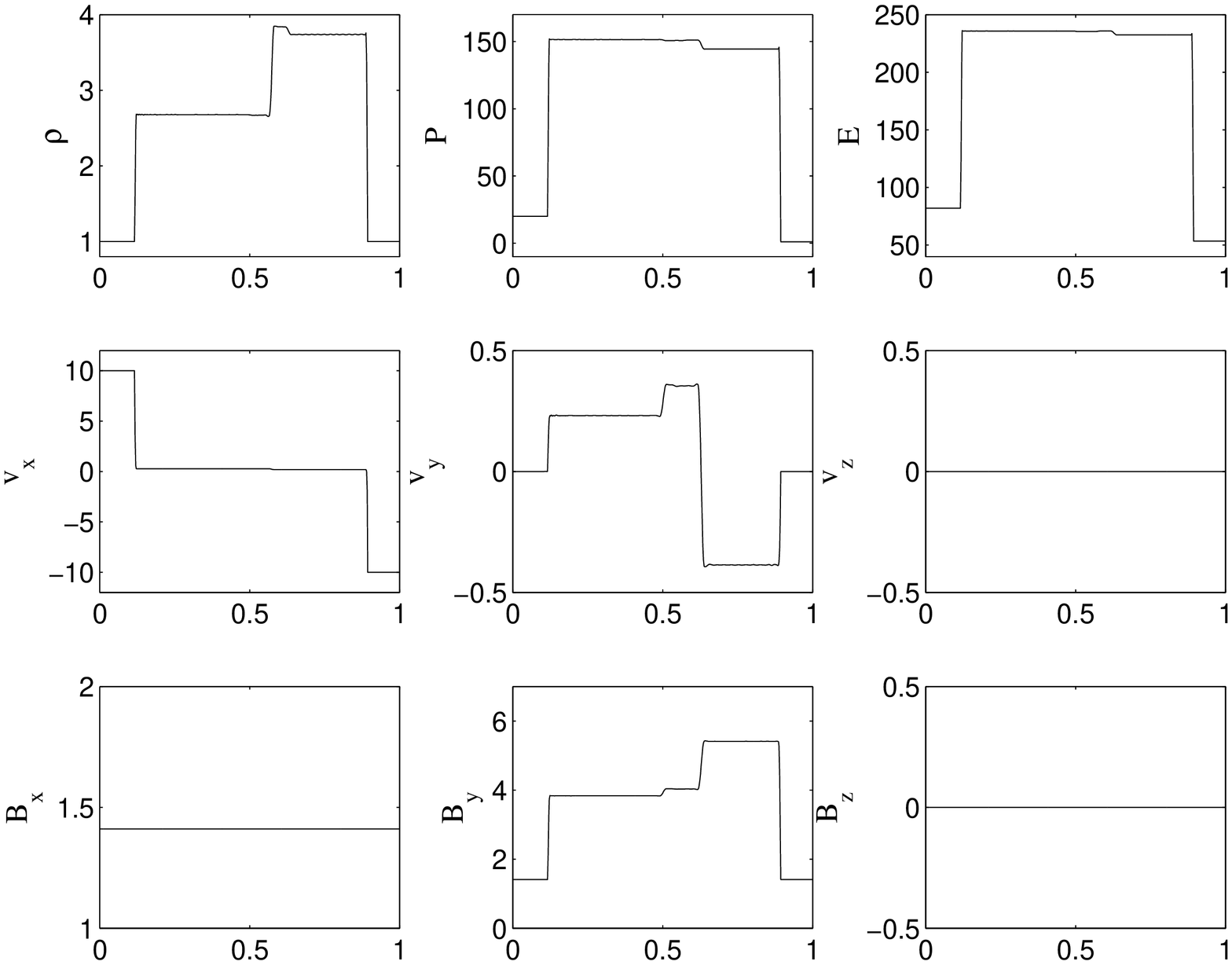}
\caption{One of the \citet{ryu95} shock-tube test results using the ORION2 MHD module.  The top row, from left to right, shows the density ($\rho$), pressure ($P$), total energy density ($E_{\rm tot}$) along the shock tube, which is resolved by 512 cells.  The second row shows the three components of velocity and the bottom row shows the three components of magnetic field.  In these panels, from left to right, we can see the fast shock, slow rarefaction (apparent in the plots of $v_y$ and $B_y$), contact discontinuity, slow shock, and fast shock.
\label{rjf}}
\end{figure}

\citet{ryu95} developed a suite of shock-tube tests that are commonly used for testing MHD algorithms.  In Figure \ref{rjf}, we present the results of just one of the tests with the setting of initial conditions $(\rho,v_x,v_y,v_z,B_x,B_y,B_z,P) = (1,10,0,0,5/(4\pi)^{1/2},5/(4\pi)^{1/2},$
$0,20)$ on the left and $(1,-10,0,0,5/(4\pi)^{1/2},5/(4\pi)^{1/2},0,1)$ on the right of the contact discontinuity; here $\rho$ is the density, $v$ the velocity and $P$ the gas pressure.  The contact discontinuity is located at the middle of the shock tube, and we set the adiabatic index of the gas to be $\gamma = 5/3$.  The length of the shock tube is 1 and is resolved by 512 cells along the $x$-direction.  The initial condition on the velocity is a colliding flow with a magnetic field at an angle of $45^\circ$ in the $x-y$ plane.  The results at a time 0.08 in code units (corresponding to 0.46 sound crossing times at the left side of the shock tube initially) are shown in Figure \ref{rjf}, and they agree with the magnitudes and locations of the shocks found by \citet{ryu95} to almost within the thickness of the lines.

\begin{figure}
\includegraphics[scale=0.7,angle=-90]{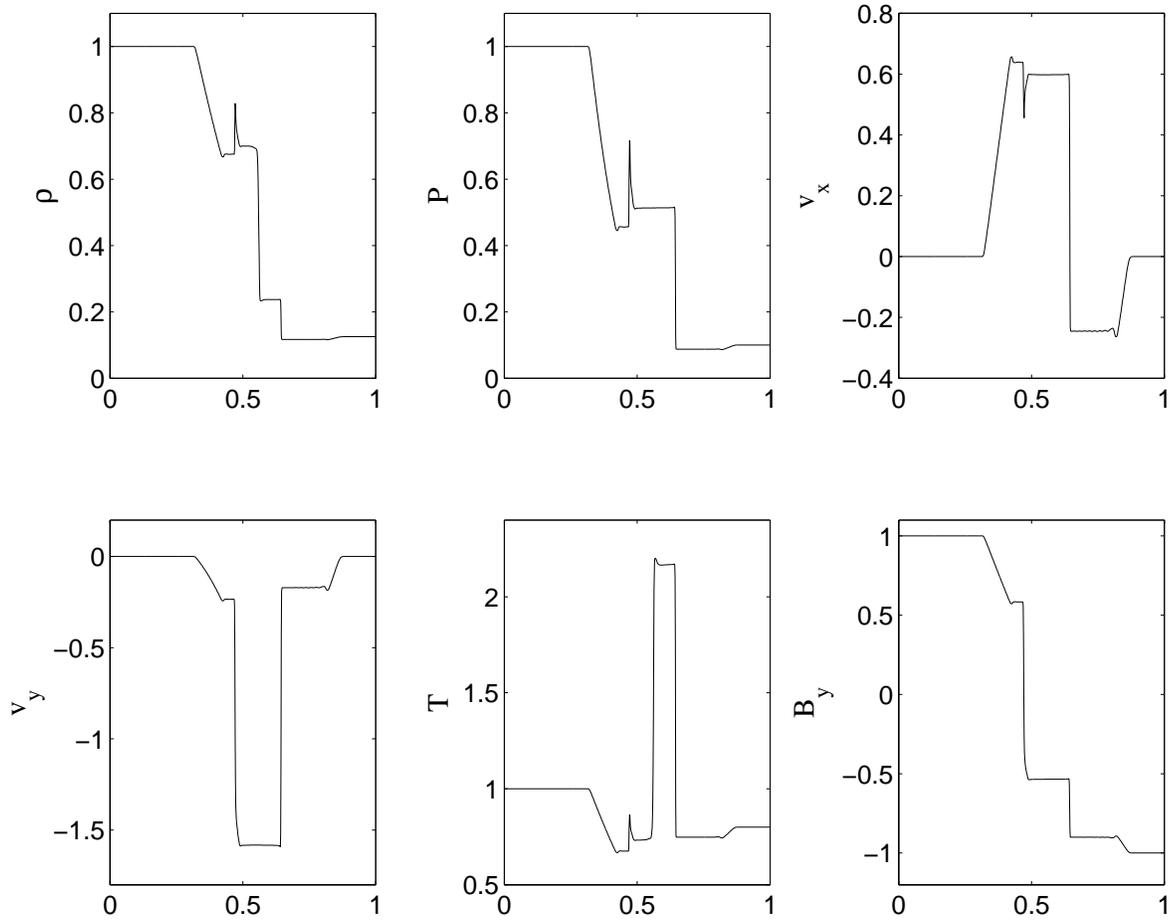}
\caption{\citet{bri88} shock tube test results.  The top row, from left to right, shows the density ($\rho$), pressure ($P$), and velocity along the shock tube ($v_x$).  The second row shows the y-component of the velocity ($v_y$), the temperature ($T$), and the y-component of the magnetic field ($B_y$).  The shock tube is resolved by 800 cells.  A compound wave composed of an \alfven and a slow wave is seen as the spike near the middle of the figure.
\label{briowu}}
\end{figure}

In Figure \ref{briowu}, we show the results of another commonly used shock-tube problem, that of \citet{bri88} .  The initial conditions of this test are $(\rho,v_x,v_y,v_z,B_x,B_y,B_z,P) = (1,0,0,0,0.75,$
$1,0,1)$ on the left and $(0.125,0,0,0,0.75,-1,0,0.1)$ on the right; here $P$ is the gas pressure.  The contact discontinuity is located at the middle of the shock tube and  $\gamma = 2$.  The length of the shock tube is 1 and is resolved by 800 cells along the $x$-direction.  The gas has no movement initially.  The $y$-component of the magnetic field changes sign at the contact discontinuity and has a pressure jump of 10.  Figure \ref{briowu} shows the results at a time of 0.1 (corresponding to 0.14 sound crossing times at the left side of the shock tube initially).  We can see the compound wave, which is composed of an \alfven wave and a slow wave, in the figure.

\subsection{Field-Loop Advection}
\label{fieldloop}

\begin{figure}
\includegraphics[scale=0.5]{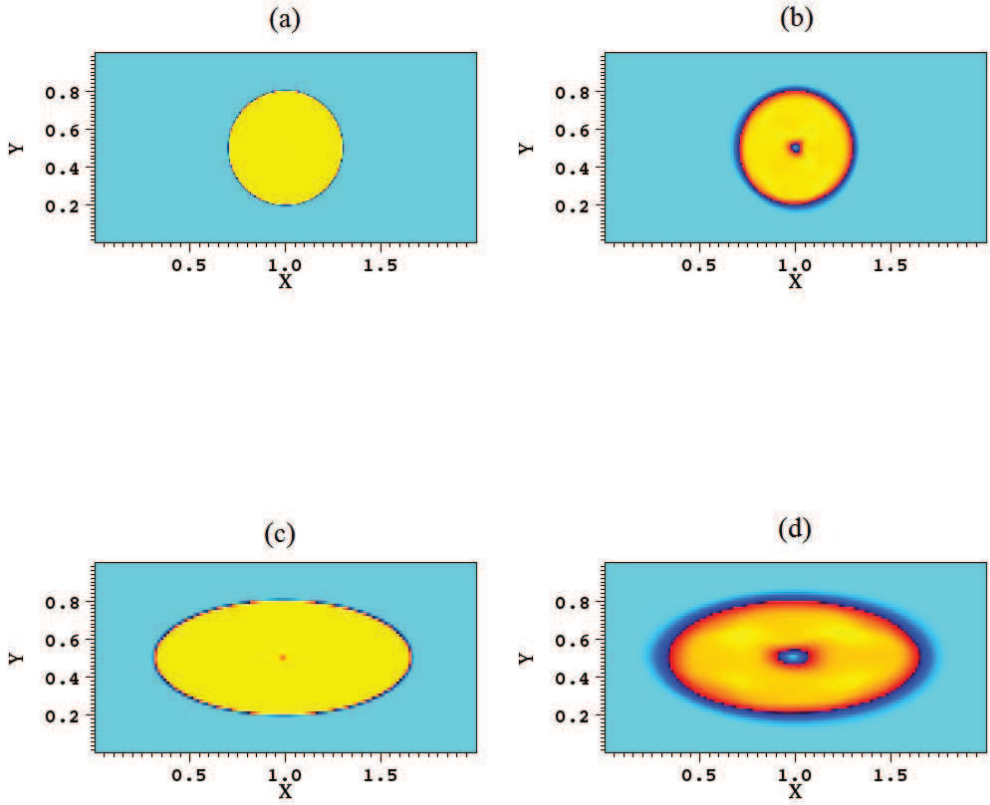}
\caption{Vertical magnetic field-loop advection test on a unigrid with resolution $N = 64$ (a) at the beginning of the advection and (b) after 2 cycles of advection.  The inclined magnetic field loop, which is rotated by 63.4\degr~about the $y$-axis from the vertical, (c) at the beginning of the advection and (d) after 2 cycles of advection.  Both field loops are advected with velocity (2,1,1).  All four panels are at z=0.5.  The field loops maintain their shapes accurately after 2 cycles of advection. (A color version of this figure is available in the online journal.)
\label{loop1}}
\end{figure}

\begin{figure}
\includegraphics[scale=0.7,angle=-90]{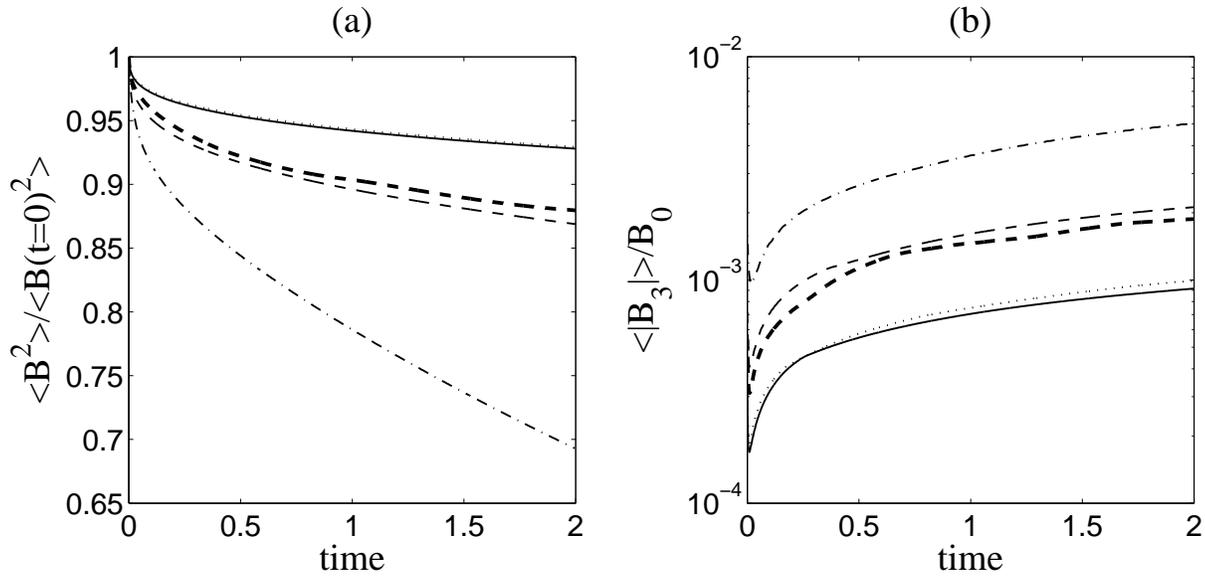}
\caption{Time evolution of (a) the mean-squared magnetic field, $\avg{B^2}$, normalized by the initial value at time = 0 and (b) the normalized error, $\avg{|B_3|}/B_0$, for the inclined magnetic field-loop advection tests.  The three unigrid tests are shown as thin dot-dashed ($N = 32$), thin dashed ($N = 64$), and thin solid ($N = 128$) curves.  The dotted curve (which nearly overlaps the $N = 128$ unigrid test result) is from the advection test with AMR at $N = 64$ resolution  
on the base level and 1 level of refinement (see Figure \ref{loop4}).  The thick dashed curve is from the advection test with Fixed Mesh Refinement (FMR) at $N = 64$ resolution at the base grid and 1 level of refinement on a fixed fine grid smaller than the base grid (see Figure \ref{loop5}). Both the AMR and the FMR tests use a refinement ratio $n_{\rm ref} = 2$.  See \S \ref{fieldloop} for discussion of the tests.
\label{loop2}}
\end{figure}

\begin{figure}
\includegraphics[scale=0.5,angle=0]{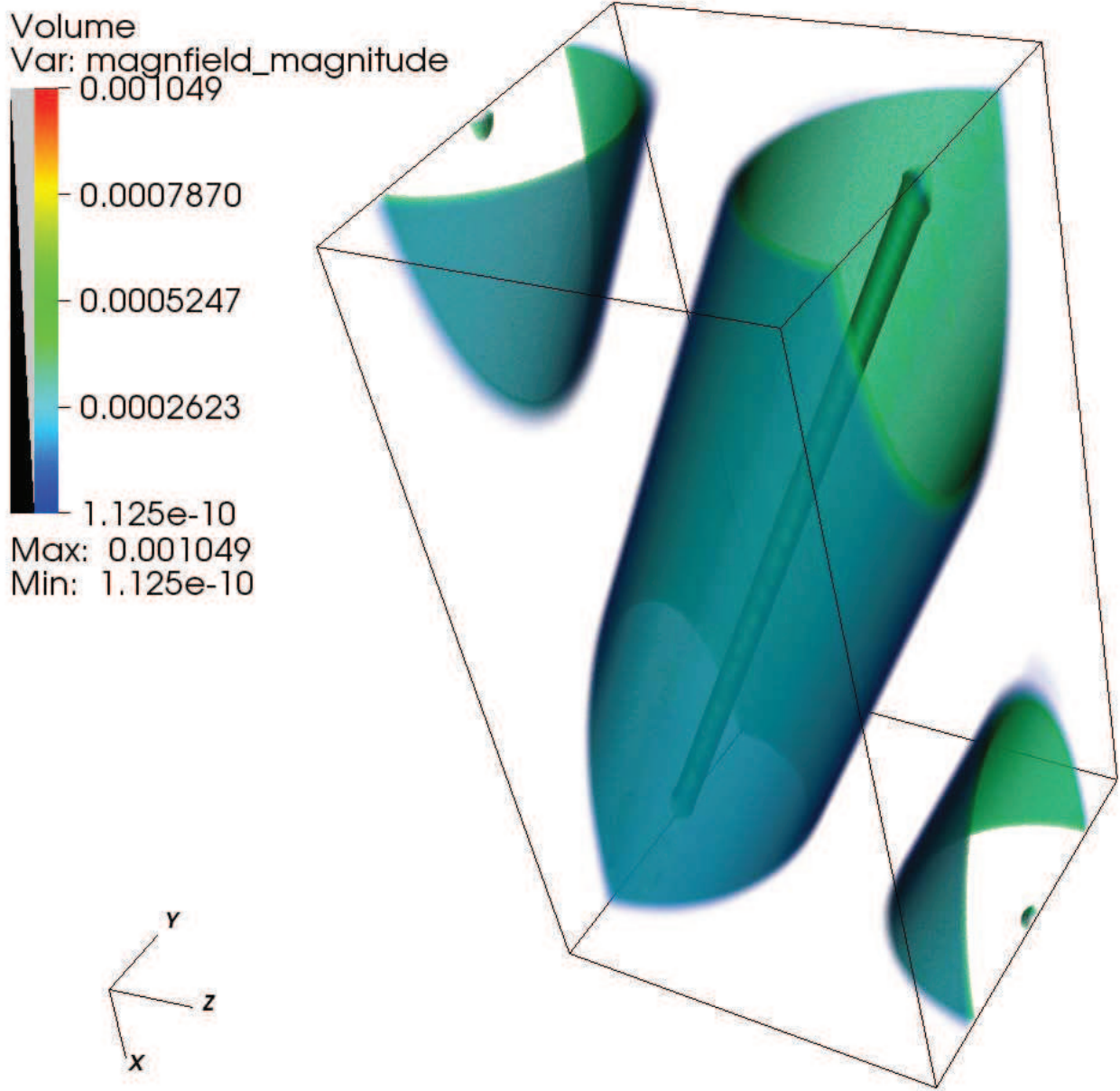}
\caption{Volume rendering of the inclined field loop from the 3D $N=64$ unigrid model after two cycles of advection along the box diagonal. (A color version of this figure is available in the online journal.)
\label{loop3}}
\end{figure}

Since \citet{gar05} first suggested using the advection of a magnetic field loop to show the difference between operator split and unsplit schemes, field-loop advection has become a standard test for MHD algorithms.  Our setup of the 3D
test is exactly the same as in \citet{gar08}: the magnetic field loop is created inside a rectangular box of size (2,1,1) resolved on a $2N \times N \times N$ grid with periodic boundaries.  The loop of magnetic field is generated from a vector potential on a coordinate system ($x_1,x_2,x_3$), with $A_1 = A_2 = 0$ and
\begin{eqnarray}
A_3 = \left\{
\begin{array}{cl}
B_0(R-r) & ~~~{\rm for} ~r \leq R,\\
0        & ~~~{\rm for} ~r > R,
\end{array} \right.
\end{eqnarray}
where $B_0 = 10^{-3}$, $r = \sqrt{x_1^2+x_2^2}$, and the size of the loop is $R = 0.3$.  The computational coordinate system ($x,y,z$) is transformed to ($x_1,x_2,x_3$) by
\begin{eqnarray}
&x_1& = (-2x+z)/\sqrt{5}, \nonumber\\
&x_2& = y,\\
&x_3& = (x+2z)/\sqrt{5}, \nonumber
\end{eqnarray}
corresponding to a rotation about the $y$-axis.  The density ($\rho = 1$) and pressure ($P = 1$) are uniform and the whole region is advected with a velocity (2,1,1).  Therefore, after one unit of time, a loop starting in the middle of the rectangular region will travel across the 3D diagonal of the box and return back to the initial position.  The advection continues for 2 cycles and the images of the magnetic field loop at the beginning and the end of this test are shown in Figure \ref{loop1}.  The top part of the figure shows the simpler case in which 
the field loop is aligned with the $z$-axis, which is similar to a 2D advection test.  The loops diffuse slightly but maintain their shapes nicely after 2 cycles of advection.  The time evolution of the volume mean $\delta B^2 / B^2$ is similar to that of the 3D inclined field-loop test, and $B_z$ remains zero to the machine accuracy.  In the rest of this section, we focus on the 3D inclined field-loop test results.  We have performed the inclined field-loop test 3 times on a single level grid (unigrid) with 3 different resolutions: $N = 32,\,64$ and 128, as in \citet{gar08}.  No AMR is used.  The time evolution of the mean square field, $\langle B^2\rangle$, normalized by the initial value, $\langle B_i^2\rangle$, of the whole advection sequence is shown in Figure \ref{loop2}a as the three thin curves.  The results are similar to those of \citet{gar08}, who used a second-order PLM reconstruction scheme, and to those of \citet{fro06}, who used a second-order TVD scheme for the Ramses code.  \citet{gar08} pointed out that with the axis of the field loop aligned along an inclined direction with respect to the computational grid, preserving $B_3$ to be zero is non-trivial.  The time evolution of the normalized error, $<|B_3|>/B_0$, of the 3 tests is shown in Figure \ref{loop2}b; the error is slightly smaller than that in \citet{gar08}.
Figure \ref{loop3} shows the volume rendering of the inclined field loop from the $N = 64$ unigrid model after 2-cycles of advection along the 3D box diagonal.

\begin{figure}
\includegraphics[scale=0.5]{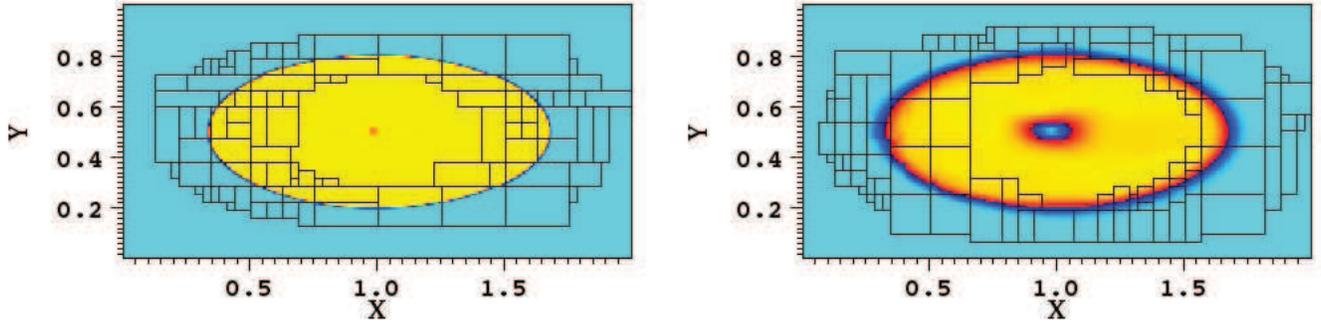}
\caption{Inclined magnetic field-loop advection test with AMR at the beginning of the advection (left panel) and after 2 cycles of advection (right panel).  The blocks surrounding the loop show the location of the fine grids.  Although the base grid has a resolution of $N = 64$, the loop is always refined at a resolution of $N = 128$.  The time evolution of the normalized $\avg{B^2}$ and $\avg{|B_3|}$ is almost the same as that of a unigrid $N = 128$ test (Fig. \ref{loop2}). (A color version of this figure is available in the online journal.)
\label{loop4}}
\end{figure}

We next investigate how AMR affects this test.  In Figure \ref{loop4}, we show an AMR advection test with a base grid of resolution $N = 64$ and one level of refinement, with a refinement ratio $n_{\rm ref} = 2$.  The refinement criterion is on the jump in the normalized magnetic pressure ($\delta B^2 / B^2$), and the refinement threshold is 2.8.  With this refinement criterion, the entire field loop is covered by the fine grid.  The level 1 fine grids continually move with the loop.  After 2 cycles, the loop maintains its shape as in the advection test on a unigrid with a resolution of $N = 128$.  The dotted curve in Figure \ref{loop2}a almost exactly coincides with the thin solid curve.  The normalized error $\langle |B_3|\rangle/B_0$ is also close to that of $N = 128$ test.

\begin{figure}
\includegraphics[scale=0.5]{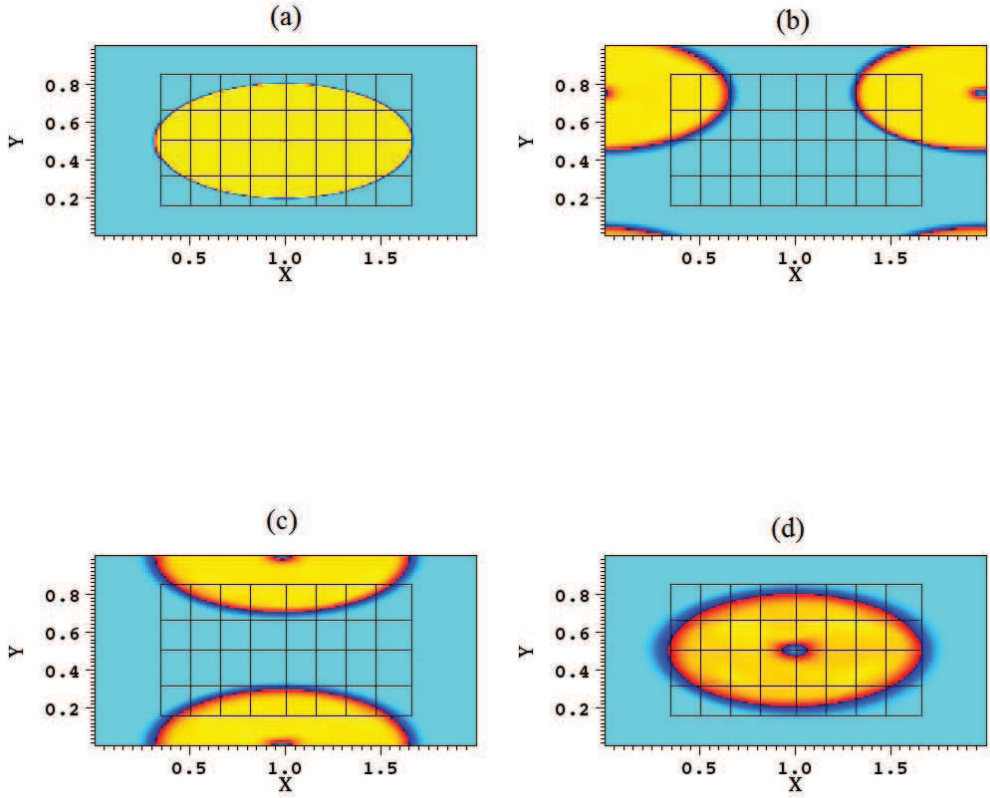}
\caption{Inclined magnetic field-loop advection test with fixed mesh refinement at (a) time = 0, (b) time = 0.25, (c) time = 0.5, and (d) time = 2.0.  The base grid resolution is $N = 64$ and the fine grid is equivalent to $N = 128$.  When the field loop crosses the coarse-fine grid boundary, the loop diffuses to the resolution of $N = 64$ but maintains its shape accurately after 2 cycles of advection.  See Figure \ref{loop2} for the time evolution of the normalized $\avg{B^2}$ and $\avg{|B_3|}$ for this fixed mesh test. (A color version of this figure is available in the online journal.)
\label{loop5}}
\end{figure}

In Figure \ref{loop5}, we show an advection test using fixed mesh refinement (FMR) as opposed to adaptive mesh refinement (AMR).  The level 1 fixed fine grid, which also has $n_{\rm ref} = 2$, is smaller than the base grid in space, and therefore the loop passes through the coarse-fine grid boundary during the advection.  The loop maintains its shape nicely after 2 cycles of advection passing in and out of the fine grid, but it diffuses slightly after the first passage through the coarse/fine boundary.  The normalized values of $\langle B^2\rangle$ and 
$\langle |B_3|\rangle$  are shown in Figure \ref{loop2}a and \ref{loop2}b
(thick dashed curve).  Note that this test also highlights the need for effective refinement criteria and for adaptive refinement that follows the solution, since this test only results in comparable accuracy to the uniform coarse mesh.  In the unigrid, AMR and FMR tests, $\divb$ vanishes to order $10^{-16}$ (machine accuracy).

\section{Supersonic Isothermal MHD Turbulence with AMR}
\label{sec:mhdturb}

\subsection{Simulation Model Parameters}
In this section, we investigate
the effects of AMR on ideal MHD simulations of isothermal, supersonic turbulence in a strong magnetic field.  We discuss only the effects of AMR on the velocity power spectrum, the density PDF, and the turbulent dissipation rate.  We present the results of 10 simulations, all with an rms 3D sonic Mach number $\mrms = 10$ and an initial value of the plasma $\beta$ paramater $\beta_0 = 0.1$; the corresponding initiial \alfven Mach number is $\mao=\surd 5$.  Periodic boundary conditions are used.  We implemented an algorithm for driving the turbulence with AMR using the recipe discussed in \citet{mac99}.  The high value of the sonic Mach number and the low value of $\beta_0$ have proven to be very challenging in the past since the simulation can become unstable within or soon after one dynamical crossing time, especially with AMR.  We define the dynamical crossing time as the length of the box divided by the rms velocity of the turbulent box.  The driving pattern we use is the same for all 10 models to facilitate direct comparison.  The system is continuously driven at the largest scales, $k = 1 \sim 2$, so as to maintain $\mrms=10$ for 3 dynamical crossing times (i.e., 0.3 sound crossing times).  We analyze the data only after the first crossing time so as to allow the system to relax to a steady state.  There are a total of 50 data files written out from each model in the last two dynamical crossing times.

We have explored different interpolation schemes and Riemann solvers in order to determine a combination of these algorithms which provides accuracy and stability for driven turbulence over several dynamical crossing times.  Here we describe the combination of algorithms we adopted for our tests:
\begin{itemize}
\item[1)]
The second-order accurate in space, piecewise total variation diminishing (TVD) linear interpolation scheme described in 
\citet{tor99}.
\item[2)]
The multi-dimensional shock flattening strategy developed 
by \citet{mig05}, in which the interpolation reverts to the minmod limiter and the fluxes are computed using the HLL solver when a strong shock is detected.
This provides additional dissipation in the proximity of a strong shock 
so as to guarantee positivity of the pressure.
\item[3)]
The harmonic mean limiter of \citet{vanl74}.
\item[4)]
With the CT scheme, the electromotive force (EMF) is computed at the zone edges using a two dimensional Riemann solver based on a four-state HLL flux function \citep{lon04,mig07}.
\item[5)]
The simple three-state HLLD approximate Riemann solver for the isothermal case 
described by \citet{mig05}.  The absence of the entropy mode in the isothermal case leads to a different formulation based on a three-state representation rather than the four-state representation of \citet{miy05}.  The MHD module can also handle a non-isothermal ideal gas, but we do not include tests with non-isothermal turbulence here.
\item[6)]
The characteristic tracing scheme of \citet{col84}.
\end{itemize}
\noindent

We have tried the Roe solver \citep{roe86}, which is an approximate linear Riemann solver, with the above combination, and it is equally stable for long-duration driven turbulence simulations.  The standard tests in \citet{mig07} show that the Roe and HLLD solvers yield comparable accuracy, but the HLLD solver is faster.  Our tests lead to the same conclusion, so we use the HLLD solver for all the driven MHD turbulence simulations in this investigation.

To push this stable scheme further, we carried out two simulations on a $128^3$ base grid with 2 levels of refinement for 3 dynamical crossing times: (1) $\mrms = 10$, plasma $\beta_0 = 0.02$; and (2) $\mrms = 17.32$, plasma $\beta_0 = 0.00667$. We found that the former test is stable using a CFL number of 0.4 and the latter  test is stable using a CFL number of 0.35.  The results for the second test (Model 11) are shown with the other 10 models for reference, even though the initial conditions of this test are different.  All the standard tests shown in \S\ref{tests} use the above combination of algorithms and demonstrate the accuracy resulting from this choice.  However, it must be borne in mind that our choice of Riemann solver, reconstruction, limiter, and EMF averaging schemes is probably not unique, since we have tested only a small combination of all existing solvers and reconstruction schemes.

We have also tried the third-order accurate Piecewise-Parabolic-Method (PPM) of \citet{col84} as the reconstruction scheme, but the simulation becomes unstable 
within or soon after one dynamical crossing time, regardless of what limiter
is used (even with the most diffusive minmod limiter).  We conclude that
special treatments will be required to use higher-order interpolation schemes for simulations of high Mach number, strong magnetic field turbulence.

\subsection{Refinement Criteria}

\citet{kri06} have proposed two refinement criteria for hydrodynamic turbulence, one based on the jump in pressure and one on the norm of the velocity gradient matrix.  We have used this as a guideline in setting our refinement criteria for MHD turbulence.  For the first criterion, we replace the thermal pressure $P$ by the sum of the thermal and magnetic pressures, $P_{\rm tot}=P+B^2/8\pi$; cells are tagged for refinement if $\Delta P_{\rm tot}/P_{\rm tot}$ exceeds a pre-determined threshold.  For the second criterion we use the \citet{kri06} refinement criterion for strong shear, which is determined by computing the norm of the velocity gradient matrix $\Vert\partial_i v_j \Vert$ without the contribution from the diagonal elements.  The norm is then normalized by $c_s/\Delta x$ and is tagged for refinement at a threshold that is the same as that for the total pressure jump.
\citet{kri06} found that AMR results agreed well with unigrid results at the maximum resolution of the AMR for a pressure jump threshold of 2.  With the inclusion of the shear velocity refinement criterion, they found that the pressure jump threshold could be raised to 3.  

We have carried out a series of tests to investigate the sensitivity to the refinement threshold.  Unlike \citet{kri06}, we find that a refinement threshold $\Delta P_{\rm tot} / P_{\rm tot} = 2$ causes the base level to be completely refined in all our AMR models, most likely because of the inclusion of magnetic pressure.  \citet{kri06} use an AMR refinement ratio $n_{\rm ref}^{\ell-1}= 4$.
For example, their $1024^3$ AMR run has a base grid of $256^3$  with one level of refinement.  All of our tests use $n_{\rm ref}^{\ell-1} = 2$, except for one with $n_{\rm ref}^0 = 4$ for comparison.  With a refinement ratio of 2, two levels of refinement are needed to achieve a maximum resolution of $1024^3$ with a $256^3$ base grid. The fraction of the volume covered by the first level of refinement (i.e., the ``coverage") is independent of the refinement ratio. Thus, the coverage at level 1 with a refinement ratio of 2 (maximum refinement equivalent to $512^3$) is the same as that at level 1 with a refinement ratio of 4 ($1024^3$), but the number of cells is 8 times larger in the latter case.  The coverage at level 2 with a refinement ratio of 2 ($1024^3$) is smaller than that of level 1 with a refinement ratio of 4, and the number of cells is correspondingly smaller.

A simple way of characterizing the refinement of a multilevel AMR calculation is the 
volume-averaged resolution,
\beq
\ravg \equiv \sum_{i} R_i V_i.
\label{eq:re}
\eeq
where $R_i$ is the 1D resolution of level $i$ and  $V_i$ is the fractional
volume coverage for level $i$ excluding higher levels.
For example, for Model 4, the volume coverage of levels 0, 1 and 2 is $V_i=0.35,\, 0.48,\, 0.17$, respectively (the total level 1 coverage is 0.65 but $V_1$ excludes the volume refined at level 2).  For this model, $\ravg=128\times0.35 + 256\times0.48 + 512\times0.17 = 255$,
which is very close to that of a $256^3$ unigrid model. 

We have performed a large number of ideal MHD turbulent box experiments using different base grid sizes ($128^3$ and $256^3$), different refinement criteria, different minimum block sizes, and different refinement ratios.  We also have two unigrid turbulence models, at $128^3$ and $512^3$, for comparison.  The unigrid $512^3$ model will serve as the reference for all AMR models presented in this paper.

\subsection{Power Spectrum}
\label{power}

\begin{figure}
\includegraphics[scale=0.7,angle=-90]{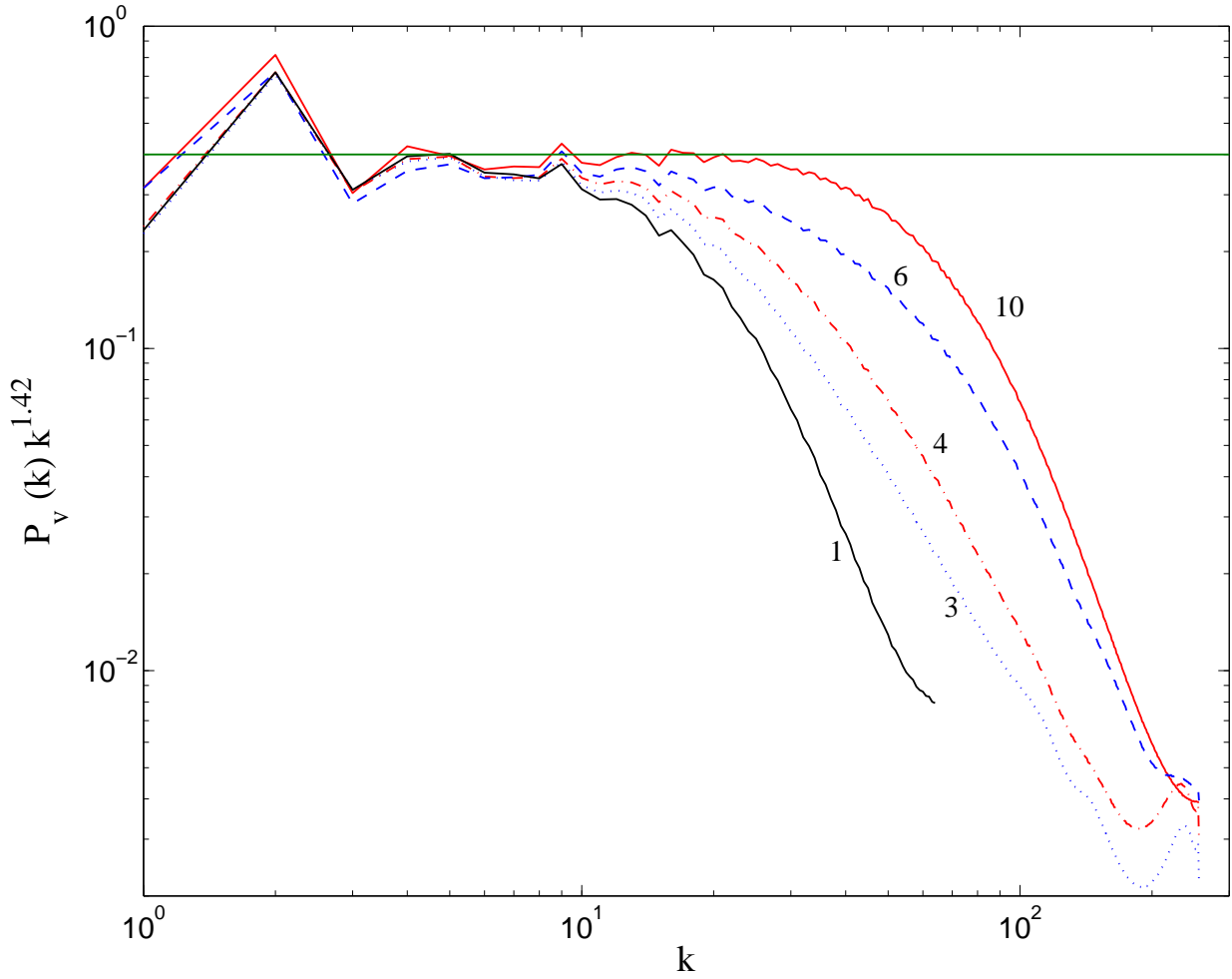}
\caption{Compensated velocity power spectra of models 1, 3, 4, 6, and 10.  The spectra are compensated by $k^{1.42}$, the spectral index of the unigrid $512^3$ Model 10.  Models 3, 4, and 6 are simulations with AMR.  Models 1 and 10 have just the base grid at $128^3$ and $512^3$, respectively.  The spectra are labeled by their model numbers.  The horizontal line is provided to aid comparison of the inertial ranges of all the models.  See \S \ref{power} for a discussion of how refinement coverage affects the inertial range. (A color version of this figure is available in the online journal.)
\label{ps}}
\end{figure}

In this section, we study how changes in resolution and refinement criteria affect the velocity power spectrum, $P_v(k) \propto k^{-n}$, in terms of the power index,  $n$, and the extent of the inertial range, $\kmax$.  We fit the power spectra obtained from the 50 data dumps between 1 and 3 dynamical crossing times between $k=4$ (to avoid the effects of driving) up to a value of $k$ that increases by unity at each iteration.  
All the fitting results and the plots shown in section \ref{sec:mhdturb} are time-averaged results from the 50 data dumps over these two dynamical crossing times.
(Since the turbulence remains correlated for about one dynamical crossing time \citep[e.g.][]{li08}, we assign an error to the mean value of $P(k)$ from the 50 data dumps equal to the standard deviation divided by $\surd 3$.)
As more points are added, the uncertainty in the slope decreases, and correspondingly so does the reduced $\chi^2$ of the fit. However, the reduced $\chi^2$ begins to increase when the power spectrum turns over due to numerical dissipation; we define the point at which $\chi^2$ is a minimum as the upper end of the inertial range, $\kmax$. In order to overcome the possibility that noise in the data could artificially lower $\kmax$, we omit the point at $\kmax+1$ that led to the increase in $\chi^2$ and evaluate the reduced $\chi^2$ of the fit including the point at $\kmax+2$; if the value of the reduced $\chi^2$ is less than the previous minimum,  we set $\kmax=\kmax(\mbox{previous})+2$ and proceed with the iteration. We allow for the possibility that the noise fluctuation could be up to three cells wide.  This procedure allows our estimate of the inertial range to extend beyond the bumps at $k\simeq 9$ that are apparent in the spectra in Figure \ref{ps} and are due to an artifact in the driving pattern.  This method is conservative, but it can eliminate the impact of the bottleneck effect on determining the spectral index or the size of the inertial range \citep[e.g.][]{ver07,kri07} when the inertial range available for turbulent energy transfer is small.  However, we note that our MHD turbulence tests do not suffer from the bottleneck effect.

In Table 1, we present the fitted results of the 10 ideal MHD turbulence models.  
The table includes the refinement coverage at each fine level.   For example, in Model 4, level $\ell=1$ has a volumetric coverage of 65\% of the computational domain and level $2$ has a volumetric coverage of 17\% of the computational domain
(or $17/65 \sim 26\%$ of level $1$).  In Figure \ref{ps}, the compensated power spectra of models 1, 3, 4, 6, and 10 are shown.  We summarize the results in Figure \ref{ps} and Table 1 as follows:

\begin{itemize}
\item[1)]
Models 1 and 10 are unigrid simulations and provide a basis for evaluating the AMR models. The spectral index of Model 10 is $n=1.42\pm0.02$, consistent with other strong magnetic field supersonic turbulence simulations, which show that the power spectrum is close to the Iroshnikov-Kraichnan spectrum \citep{iro63,kra65}.
The low resolution Model 1 has a steeper spectral index, $n=1.75\pm0.06$.
We repeated Model 10
(unigrid $512^3$ with an HLLD solver) using the Roe solver and obtained a
power spectrum that agrees to within the
uncertainty of fitting.
\item[2)]
Models 2 to 4 test the effect of changing the
refinement threshold for the total pressure jump from 3.25 to 2.5.  As the threshold drops, the refinement coverage increases, and the spectral index slowly approaches that of Model 10.  Correspondingly,
the inertial ranges are significantly longer (2 times) than in Model 1  and appear to converge to that of Model 10 as the refinement coverage increases.

\item[3)]
Model 5 tests the effect of shear flow refinement on the AMR calculation.  
There is no noticeable increase in the accuracy of Model 5 compared to Model 4,
presumably because the additional criterion did not significantly increase 
the average refinement. Model 8 has shear flow refinement, whereas Model 7 does not;
however, Model 8 has 2 levels of refinement compared to 1 for Model 7, so no
inference on the effect of the shear flow refinement can be drawn.

\item[4)]
Comparison of models 5 and 6, and of models 4 and 7, addresses the effects of 
refinement coverage on the overall improvement of the power spectrum.  Model 6 has only 1 level of refinement but uses a refinement ratio of 4, equivalent to full coverage of level 1 at the resolution of level 2 in Model 5, which is a 2-level model using a refinement ratio of 2.  The average refinement in Model 6 is about 60\% greater than in Model 4; correspondingly, there is a substantial improvement to the power spectrum and a modest increase in the size of the inertial range.  However, the computational time for Model 6 is about 3 times that of Model 5, a big price to pay for the improvement.  The reason for the large increase in computing time is that each cell that is refined only to level 1 in the two-level run has 8 times as many refined cells in the one-level run.  An additional test of the effects of refinement coverage is provided by comparing Model 7 to Model 4; the former uses a base grid of $256^3$ and only 1 level of refinement. Model 7
has an average refinement about 20\% greater than Model 4. For this, one gets a significant improvement in the spectral index, but only a small increase in the size of the inertial range.  The computation time for Model 7 is similar to that for Model 6.

\item[5)]
Like Models 2-4,
Models 8 and 9 address the implications of increased refinement coverage for improvements in the power spectrum.  Note that the level 2 resolution in these models is the same as that in a $1024^3$ resolution simulation.  Model 9 has only a slightly lower threshold for the pressure jump than Model 8 (2.3 vs. 2.5), but
this leads to almost a twofold increase in the average refinement. This increase in 
average refinement yields an increase in the inertial range, but has no significant effect on the spectral index. Even though the average refinement of Model 9 exceeds that in the unigrid Model 10, the latter appears to have the most accurate spectral index and the largest inertial range.  Our finding that a unigrid simulation at $512^3$ resolution is superior to an AMR simulation with a maximum refinement of $1024^3$ is consistent with the conclusion of \citet{kri06} that a large base grid is required to obtain the advantages of AMR for simulations of turbulence.

\item[6)]
Model 11 has different initial conditions from the above 10 models.  This model has $\mrms = 17.32$ and plasma $\beta_0 = 0.00667$; the corresponding \alfven Mach number is $\mao=1$.  This model is in many ways similar to Model 3 in Table 1 in terms of the power spectral index, the length of inertial range, and refinement coverage, although Model 3 has very different initial conditions ($\mrms=10$, $\beta_0=0.1$, and $\mao=\sqrt{5}$).

\end{itemize}
\noindent
From the above summary, we can see that changes in refinement criteria directly affect the average refinement, which in turn directly affects the quality of the turbulence power spectrum.

\begin{figure}
\includegraphics[scale=0.7,angle=-90]{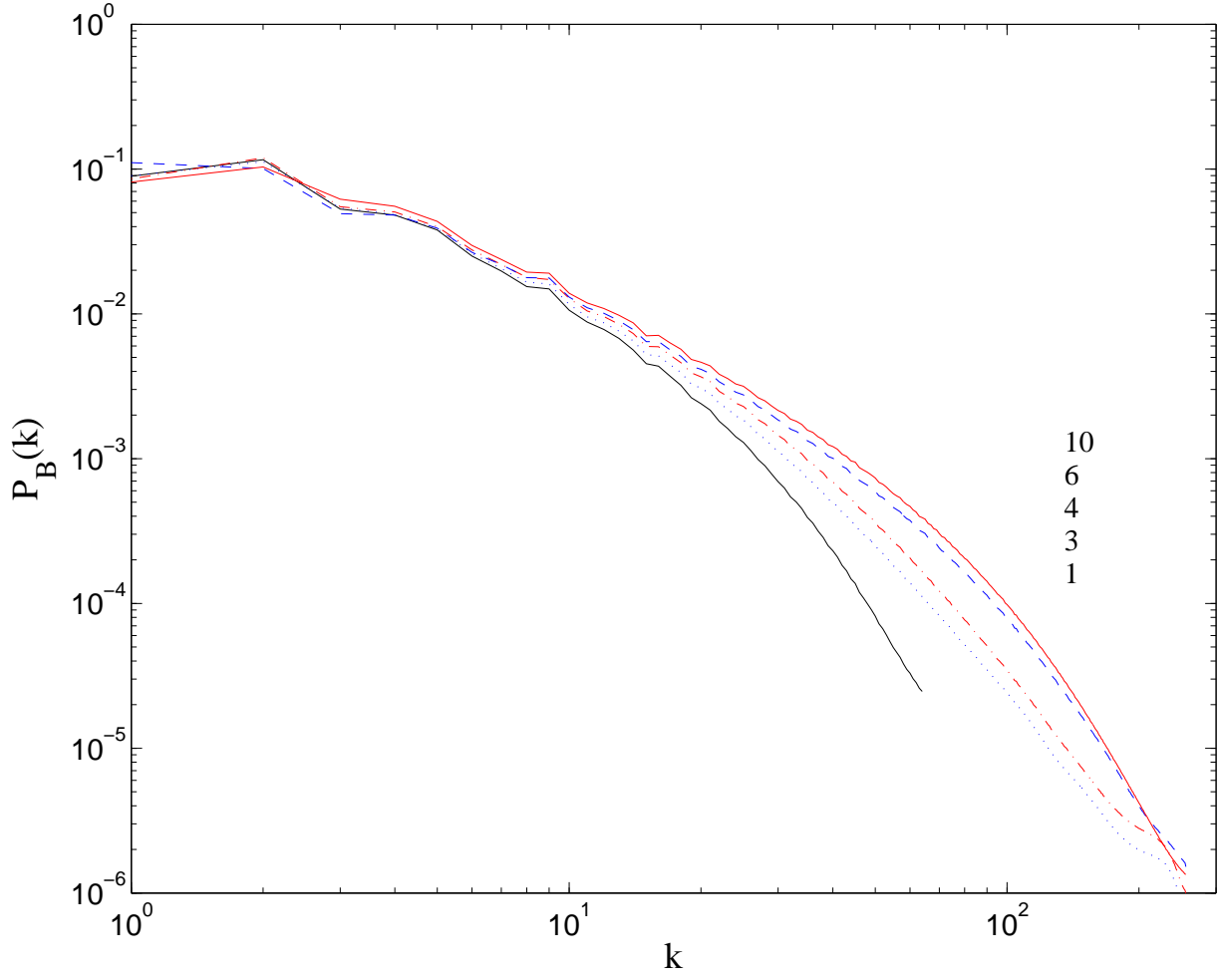}
\caption{Magnetic field power spectra of models 1, 3, 4, 6, and 10 (with model numbers shown in the same order as the curves, starting from bottom) without compensation.  The spectra show a similar convergence behavior with resolution as the velocity spectra in Figure \ref{ps}. (A color version of this figure is available in the online journal.)
\label{bps}}
\end{figure}

In Figure \ref{bps}, we show the power spectra of the magnetic field for models 1, 3, 4, 6, and 10.  The convergence behavior of the magnetic field power spectrum is similar to that of the velocity power spectra in Figure \ref{ps}.

\begin{figure}
\includegraphics[scale=0.7,angle=-90]{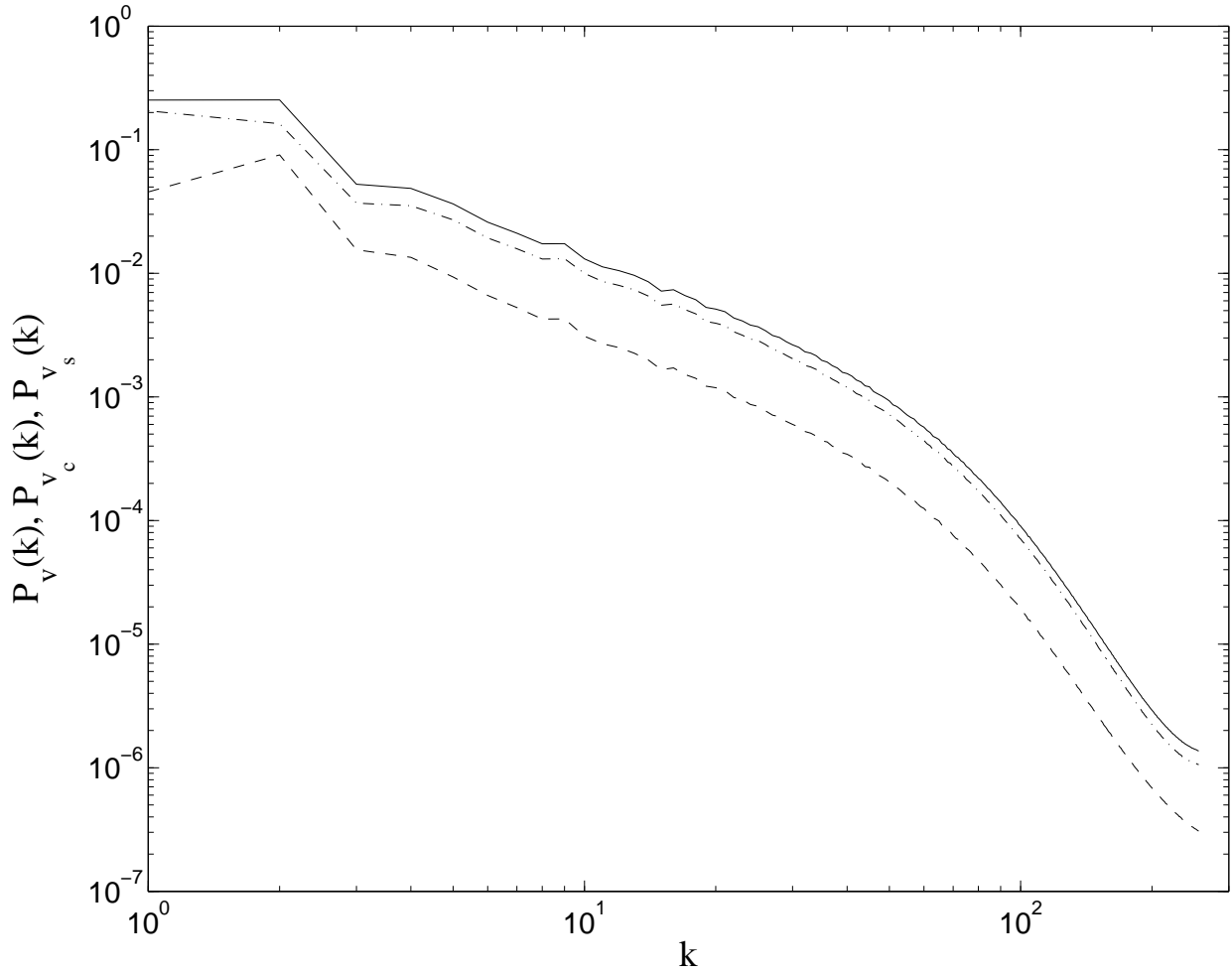}
\caption{Time-averaged velocity power spectra of Model 10.  The solenoidal $P_{v_s}(k)$ (dot-dashed) and the dilatational $P_{v_c}(k)$ (dashed) components of the velocity spectrum are also shown.  The time-averaged fraction of dilatation component $\chi_c(k) = 0.26\pm0.02$.  The solenoidal component dominates as a result of purely solenoidal driving of the turbulence.
\label{helmd}}
\end{figure}

A recent study of the effects of purely solenoidal and purely compressive turbulent driving \citep{fed10} shows that there are significant differences in the turbulence statistics between these two extreme driving models.  For example, the dispersion in the density PDF with purely compressive driving can be 3 times that resulting from purely solenoidal driving.  For purely solenoidal driving, the solenoidal component of the power spectrum will dominate the dilatational component. This is reversed for purely compressive driving.  Reality probably is somewhere in between these two extreme cases.  Our driving is purely solenoidal.  We decompose the solenoidal ($\vecnabla \cdot \vecv_s \equiv 0$) and dilatational ($\vecnabla \times \vecv_c \equiv 0$) 
components from the velocity, $\vecv = \vecv_s + \vecv_c$, by
\beqa
\vecv_c(\veck) &=& [\hat{\veck} \cdot \vecv(\veck)]\hat{\veck} \\
\vecv_s(\veck) &=& [\hat{\veck} \times \vecv(\veck)]\times \hat{\veck}
\eeqa
\citep{lem09}.  We define the fraction of the dilatational component in the velocity power spectrum as
\beq
\chi_c(k) \equiv \frac{P_{v_c}(k)}{P_{v}(k)}
\eeq
\citep{kri09a}.  Figure \ref{helmd} shows the time-averaged 
velocity power spectrum of Model 10 with the solenoidal and dilatational components.  The time-averaged $\chi_c(k) = 0.27\pm0.01$, $0.26\pm0.01$, and $0.26\pm0.02$ for Models 1, 4, and 10, respectively.  It appears that $\chi_c(k)$ is not sensitive to the resolution.  For Model 11, which has a smaller \alfven Mach number, $\chi_c(k) = 0.27\pm0.02$ is the same.  The value of $\chi_c(k)$ in an ideal MHD turbulence model with purely solenoidal forcing (\citet{kri09a}) is $\approx 1/4$, similar to our values.  
\citet{lem09} measure the dilatational component of the 
kinetic energy (i.e., the density-weighted velocity power spectrum) and find that it is
even smaller compared to the solenoidal component.

\subsection{Density PDF}

\begin{figure}
\includegraphics[scale=0.7,angle=-90]{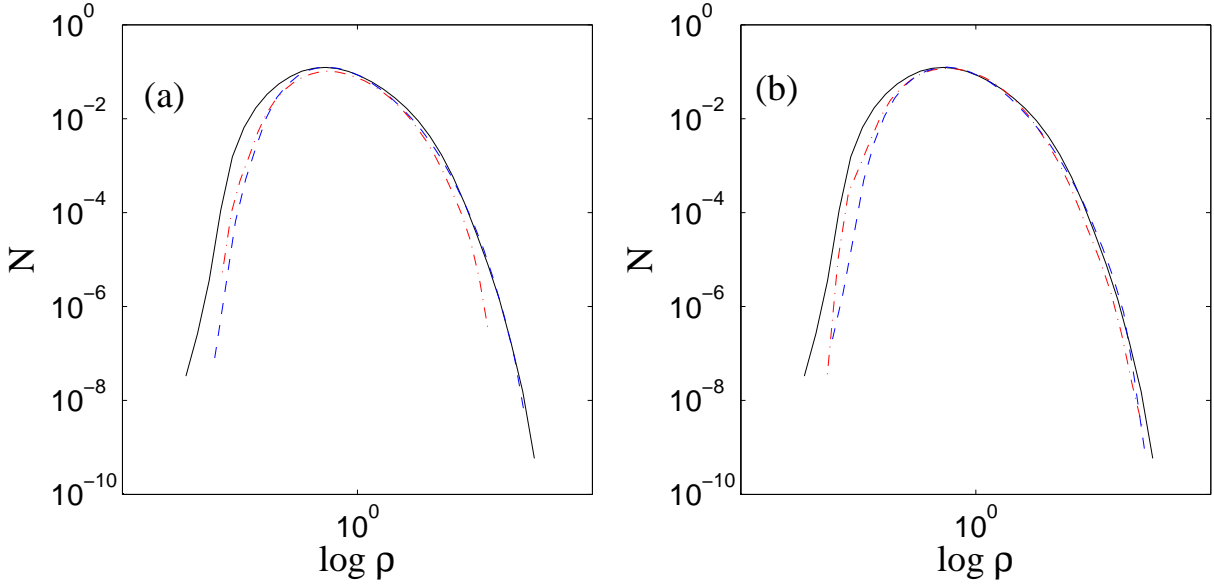}
\caption{Comparison of density PDFs.  Panel (a) shows unigrid Model 1 (dot-dash red line), AMR Model 4 (dashed blue line), and unigrid Model 10 (solid black line).  The AMR model matches the high-resolution unigrid Model 10 very well at high densities, but it is similar to the low resolution unigrid Model 1 at low densities.  Panel (b) shows AMR Model 2 (dot-dash red line), AMR Model 3 (dashed blue line), and unigrid Model 10 (solid black line).  Model 2 has a very low volume coverage at refinement level 2 and as a result the high density end deviates more from the high-resolution unigrid model than does Model 3. (A color version of this figure is available in the online journal.)
\label{pdf}}
\end{figure}

In Figure \ref{pdf}a, we show the probability density functions (PDFs) of density for models 1, 4, and 10 to demonstrate the effect of refinement on the density PDF in AMR simulations.  The density PDF of the unigrid Model 1 at $128^3$ resolution is narrower than the density PDF of the unigrid Model 10 at $512^3$ resolution.  A low resolution unigrid turbulence simulation cannot reach the highest and lowest densities attainable in a high resolution unigrid simulation.  As a result, the maximum wave speed in a low resolution simulation can be lower than in a high 
resolution simulation, since the minimum density is higher and the maximum \alfven velocity is most likely smaller.  On the other hand, since the maximum density in a low resolution simulation is smaller, the clump mass function will be reduced at high densities; this could be problematic in simulations of star formation.

AMR offers the best of both worlds: As shown in Figure \ref{pdf}a, at low densities the AMR simulation (Model 4) is close to Model 1 and therefore does not have the very high \alfven velocities that appear in the high resolution Model 10. In simulations of star formation, the loss of resolution at low densities is not important. However, Model 4 has the same maximum resolution as Model 10, and the AMR enables it to track accurately the high-density portion of the density PDF, which is critical in simulations of star formation.

In Figure \ref{pdf}b, the density PDFs of models 2 and 3 are plotted on top of
the density PDF of Model 10.  The high-density part of the PDF of Model 2
deviates more from that of Model 10 than that of Model 3  because of the very
low level 2 coverage (1.3\%) in Model 2.   Model 3 suggests that in order to
have a good match to the high density part of the Model 10 PDF, the level 2
coverage must be $\ga 10\%$ for this problem.  With the Roe solver, the PDF of a unigrid simulation extends to slightly lower densities than that for the same unigrid simulation using the HLLD solver, but the high-density parts of the PDF are almost the same.  Therefore, not only is the Roe solver slower than HLLD solver per time step, but the time step in an MHD turbulence simulation will also be smaller.

\subsection{Turbulent Energy Dissipation Rate}
\label{sec:diss}
Both hydrodynamic and MHD turbulence simulations show that turbulence decays on the order of a dynamical crossing time.  
The dissipation rate of the turbulent energy is of order $\rho \vrms^3/L$.   Specifically, we write
\beq
\dot{E} = \epsilon \frac{\rho \vrms^3}{L_{\rm int}},
\eeq
where $\vrms$ is the density-weighted rms velocity of the turbulence and the integral length scale \citep{bat53} for a compressible fluid with a magnetic field is defined by
\beq
L_{\rm int} = \frac{3\pi}{2\avg{\rho v^2+B^2/4\pi}} \int k^{-1} E_{\rm tot}(k) dk,
\label{eq:intl}
\eeq
where $E_{\rm tot}(k)$ is the total energy power spectrum, including both kinetic energy and magnetic energy.  Supersonic MHD turbulence simulations (\citet{mac99}, \citet{sto98}, and \citet{lem09}) suggest that the proportionality constant $\epsilon\sim 0.5$.  \citet{kan03} summarized the results of many pure hydrodynamic incompressible simulations and showed that, with their highest resolution ($4096^3$) simulation, $\epsilon$ converged to about 0.4.

\begin{figure}
\includegraphics[scale=0.7,angle=-90]{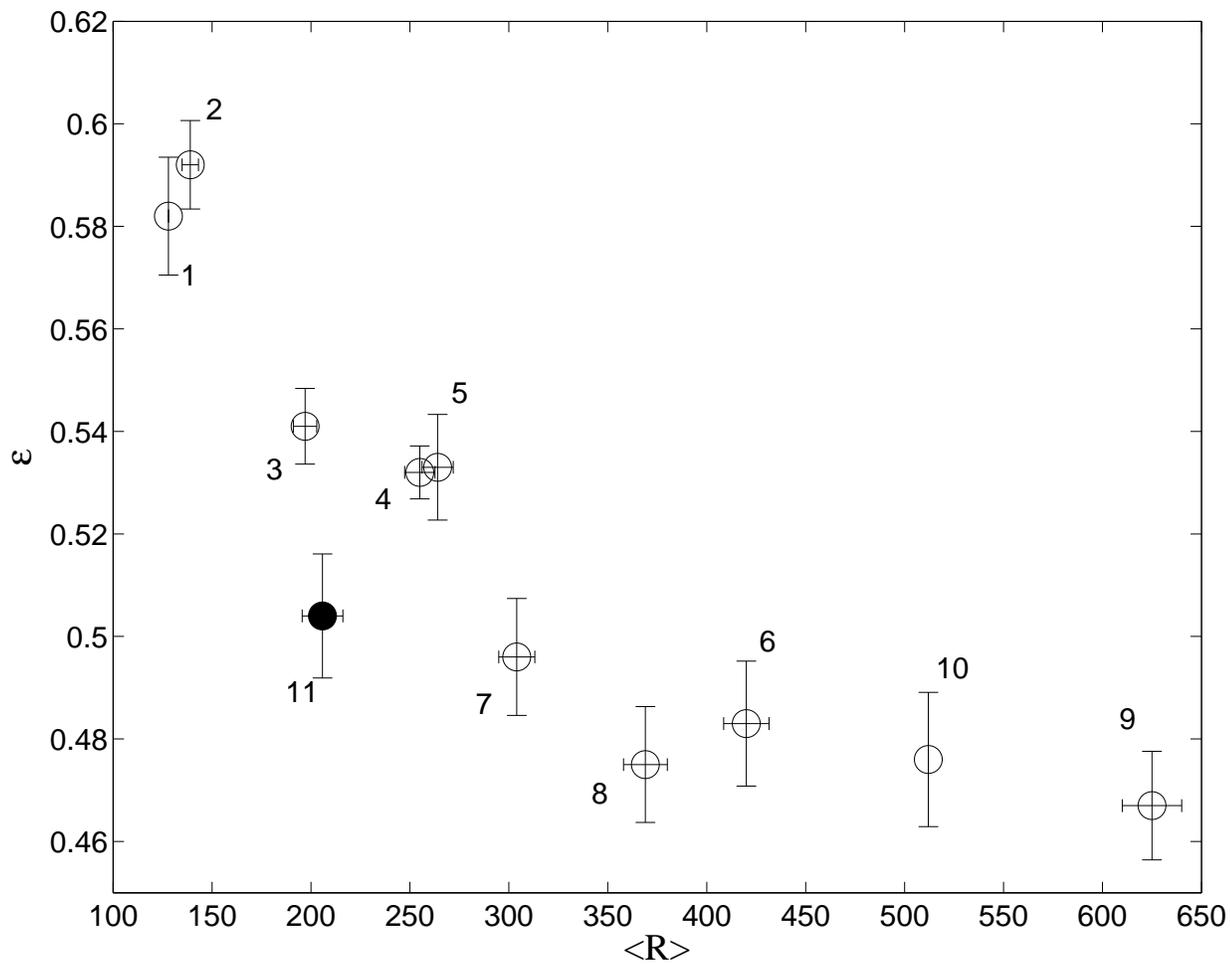}
\caption{Turbulent dissipation rate coefficient, $\epsilon$, of the 10 MHD turbulence models listed in Table 1.  For $\ravg\ga 350$, $\epsilon$ agrees with the value in the $512^3$ unigrid simulation, $0.476\pm0.011$.  Model 9 has $\ravg \sim 624$ and $\epsilon= 0.467\pm0.013$.  Model 11 (solid circle) has a somewhat smaller value of $\epsilon$ than Model 3, which has the same value of the index of the velocity power spectrum.
\label{epsilon}}
\end{figure}

The turbulent dissipation rate coefficient $\epsilon$ for the 10 models presented here is plotted as a function of the average resolution in Figure \ref{epsilon}.  The horizontal uncertainty bar is obtained from the standard deviation of the variation of the refinement coverage during the simulation, usually a few percent of the refinement coverage.  Models 1 and 10 do not have horizontal error bars since they are unigrid models.  Figure \ref{epsilon} shows that the dissipation rate is within the uncertainties of the $512^3$ unigrid simulation for $\ravg\ga 350$.  For the three AMR models with $\ravg\ga 350$, the mean $\epsilon = 0.48\pm0.02$, which agrees well with the result from the $512^3$ unigrid model, $\epsilon = 0.48\pm0.01$, and is similar to the results from other unigrid simulations of ideal MHD turbulence, $\epsilon\simeq 0.5$ \citep[e.g.][]{lem09}.  In Figure \ref{epsilon}, we also plot the value of $\epsilon$ for Model 11 (solid circle) for reference.  Although this model has the same spectral index for the velocity power spectrum as Model 3, the value of $\epsilon$ of this model is smaller.

Models 6, 8, and 9 have similar turbulent dissipation rates as Model 10, and one might think that using a lower resolution base grid with AMR would have the advantage of reaching the dissipation rate obtained from a unigrid model with higher resolution.  However, Table 1 shows that the CPU usage for these 3 models actually is higher than for the unigrid Model 10.  This is another indication of the inefficiency of AMR for turbulence studies until a much larger base grid is used.

\subsection{Turbulent Magnetic Field}

\begin{figure}
\includegraphics[scale=0.7,angle=-90]{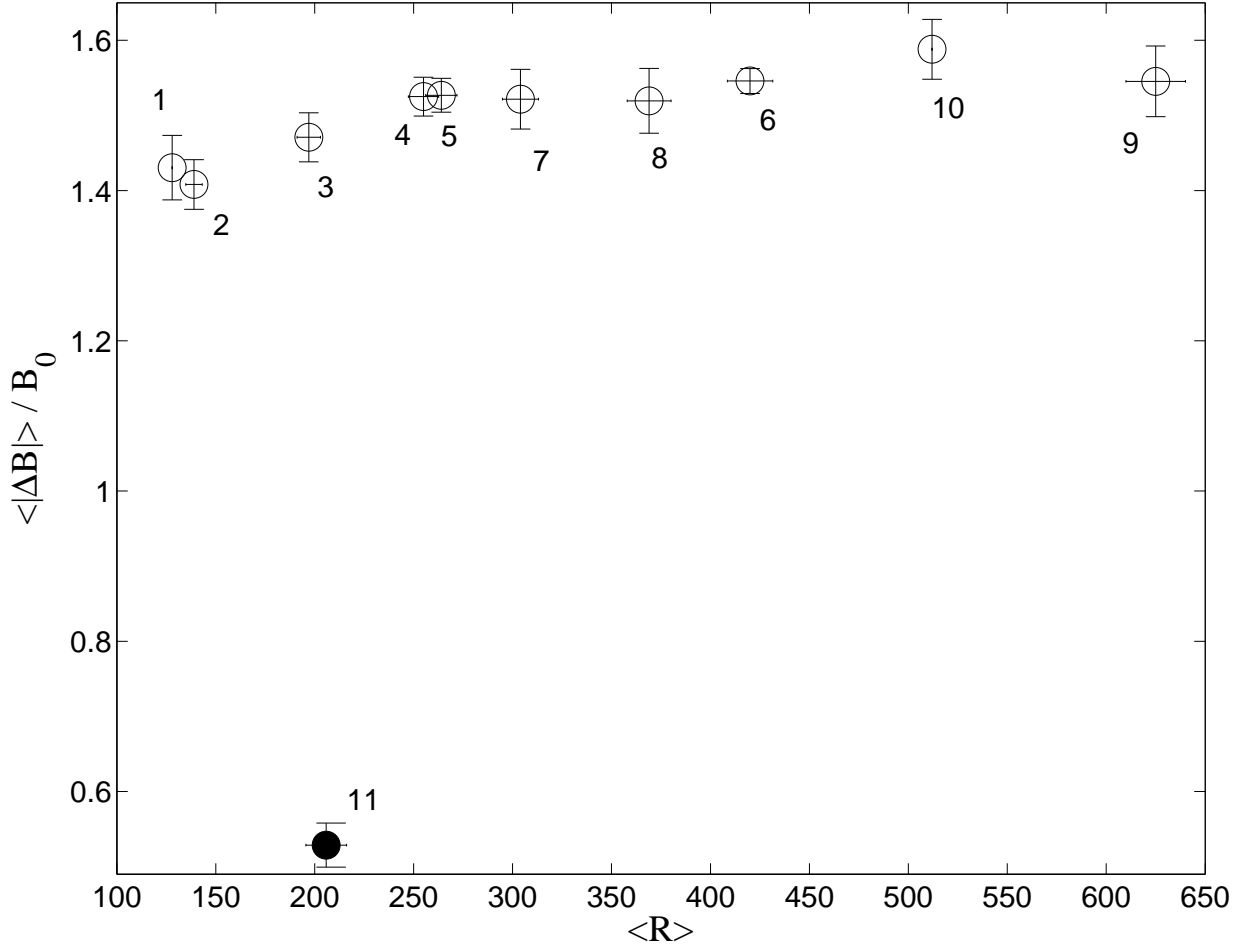}
\caption{Normalized enhancement of the magnetic field strength, $\langle | \Delta B | \rangle/B_0$, of the 11 MHD turbulence models listed in Table 1.  Model 11 ($\mao=1$; solid circle) has a smaller enhancement in the magnetic field strength than the 10 models with $\mao=\sqrt{5}$.
\label{deltaB}}
\end{figure}

For a magnetized turbulent system, the magnetic field strength will be enhanced as the result of field-line stretching.  The maximum field-line stretching will be limited by the grid resolution, which determines the numerical diffusion of the magnetic field that we want to minimize.  We plot the time-averaged change in magnetic field strength, $\langle | \Delta B | \rangle$, normalized by the initial mean field strength, $B_0$, versus the volume-averaged resolution $\ravg$ in Figure \ref{deltaB}.  Since the initial \alfven Mach number is modest ($\mao=\sqrt{5}$) for Models 1-10, the enhancement of the field strength is also modest,
$\langle | \Delta B | \rangle\sim (1.54\pm0.09) B_0$ from the mean value from Models 6 to 10, which have overlapping error bars and which have $\ravg\ga 300$.
The value of $\langle|\Delta B|\rangle/B_0 = 0.53$ for Model 11 is smaller since $\mao = 1$ is smaller.  Only systems with initially high \alfven Mach numbers
can have a turbulent magnetic field many times greater than the mean field, and simulations of such systems require much higher resolution to have converged magnetic field statistics \citep[e.g.][]{kri09a}.

\section{Conclusions and Discussions}

We have developed a robust MHD module for our new AMR code, ORION2, and have demonstrated its ability to carry out accurate long-duration AMR simulations of highly supersonic turbulent flows with strong magnetic fields ($\beta \ll 1$, $\ma\sim 1$).  Although unigrid simulations of such flows have been published (e.g., \citealp{kri09b}), to our knowledge the AMR simulations of these flows presented herein are the first to appear in the literature.  Since observations suggest that GMCs have highly supersonic flows with relatively strong magnetic fields \citep{mck07} and since AMR is essential for following gravitational collapse, this represents an important advance in our ability to study the conditions that lead to star formation.  ORION2 is able to do this because the code is sufficiently flexible that one can easily experiment with different reconstruction schemes, Riemann solvers, CT EMF averaging schemes, and limiters to find the suitable combination that we have described.

We have tested the accuracy of our code with several standard MHD tests, 
including the \citet{ryu95} shock tube test, the \citet{bri88} shock tube test, and the 3D current-loop test on unigrid, FMR, and AMR.  The results we have presented here demonstrate that the CTU+CT algorithm performs accurately and works properly within the Chombo AMR framework.  We found that the piecewise linear, spatially second-order, TVD scheme combined with a multi-dimensional shock flattening strategy developed by the PLUTO group \citep{mig07} can work with both Roe and HLLD solvers to enable stable, long-duration (3 crossing times) simulations of driven MHD turbulence with an rms Mach number of $\mrms=17.3$ and 
an initial plasma $\beta$ parameter of $\beta_0=0.0067$ on a $128^3$ base grid with 2 levels of refinement.

We examined the velocity power spectrum, the density PDF, and turbulent dissipation rate in our investigation of the effectiveness of AMR on the quality of MHD turbulence simulations.  By varying the refinement criteria, the base grid resolution, and the number of refinement levels, we find that the quality of the turbulence statistics---in particular, the spectral index of the velocity power spectrum and the extent of the inertial range---is more closely related to the average refinement coverage than to the maximum level of refinement. Analysis of the density PDFs shows that, with our refinement criteria, AMR is particularly powerful for simulations in which interest is focused on the regions of highest density: it does not capture the regions of the lowest density as well as a unigrid simulation, but it does capture the PDF of the high-density regions quite accurately \citep[e.g.][]{col11}.
Our result for the dissipation coefficient for turbulence with $\calm=10$ and $\beta_0=0.1$ is $\epsilon=0.48\pm0.01$.  This is consistent with other numerical studies of MHD turbulence with unigrid codes, which find that $\epsilon \sim 0.5$.

\acknowledgments
We would like to thank Brian Van Straalen for helping improve the scalability performance of the AMR Implementation of the MHD module of ORION2.
We also thank Christoph Fedderath, Jim Stone, Alexei Kritsuk and Tom Abel for helpful discussions on the paper.
Support for this research was provided by NASA through NASA ATP grant NNX09AK31G (RIK, CFM, and PSL), the US Department of Energy at the Lawrence Livermore National Laboratory under contract DE-AC52-07NA 27344 (RIK), 
the Lawrence Berkeley National Laboratory under contract DE-AC02-05CH11231 (DFM), and the NSF through grant AST-0908553 (CFM and RIK).  This research is also supported by an LRAC Teragrid grant of high performance computing  from the NSF and the NASA Advanced Computing (NAS) Division through a NASA ATP grant.

\clearpage

\clearpage

\clearpage

\clearpage

\clearpage

\clearpage

\clearpage

\clearpage

\clearpage

\clearpage

\clearpage

\clearpage

\clearpage

\begin{deluxetable}{ccccccccccc}
\tabletypesize{\scriptsize}
%\rotate
\tablecaption{Velocity Power Spectral Indexes, Inertial Ranges, and Refinement}
\tablecolumns{11}
\tablehead{
\colhead{Model} & \colhead{base} & \colhead{Refinement} & \colhead{Refinement} & \colhead{Shear flow} & \colhead{Spectral} & \colhead{Inertial} &
\multicolumn{2}{c}{Refinement} & \colhead{$\avg{R}$} & Normalized \\
\colhead{} & \colhead{grid} & \colhead{levels} & \colhead{threshold} & \colhead{refinement$^{\rm a}$} & \colhead{index ($n$)} & \colhead{range$^{\rm b}$} &
\multicolumn{2}{c}{coverage$^{\rm c}$ (\%)} & \colhead{} & CPU time$^{\rm d}$ \\
 & & & & & & & \colhead{$\ell=1$} & \colhead{$\ell=2$} & &
}
\startdata
1  & 128 & 0 & ...  & ... & $1.75\pm0.06$ & $4 \sim 13$ & ... & ... & 128 & 3.81(-3) \\
2  & 128 & 2 & 3.25 & no  & $1.76\pm0.07$ & $4 \sim 13$ & 6   & 1.3 & 139 & 5.17(-2) \\
3  & 128 & 2 & 2.75 & no  & $1.65\pm0.06$ & $4 \sim 14$ & 36  & 9   & 197 & 4.87(-1) \\
4  & 128 & 2 & 2.5  & no  & $1.61\pm0.03$ & $4 \sim 17$ & 65  & 17  & 255 & 6.30(-1) \\
5  & 128 & 2 & 2.5  & yes & $1.58\pm0.03$ & $4 \sim 17$ & 70  & 18  & 264 & 6.58(-1) \\
6  & 128 & 1$^{\rm e}$ & 2.5 & yes & $1.48\pm0.04$ & $4 \sim 21$ & 76 & ... & 420 & 2.46\\
7  & 256 & 1 & 2.5  & no  & $1.46\pm0.04$ & $4 \sim 18$ & 19  & ... & 304 & 2.41 \\
8  & 256 & 2 & 2.5  & yes & $1.48\pm0.04$ & $4 \sim 21$ & 31  & 6.5 & 369 & 4.71\\
9  & 256 & 2 & 2.3  & yes & $1.46\pm0.03$ & $4 \sim 25$ & 58  & 17  & 625 & 9.47 \\
10 & 512 & 0 & ...  & ... & $1.42\pm0.02$ & $4 \sim 26$ & ... & ... & 512 & 1.0 \\
11$^{\rm f}$ & 128 & 2 & 2.75 & yes & $1.65\pm0.06$ & $4 \sim 14$ & 40  & 10.4 & 206 & 4.05(-1) \\
\enddata
\tablenotetext{a}{When the shear flow refinement criterion is not included, refinement is determined by only the total pressure jump.}
\tablenotetext{b}{The inertial range extends from $k=4$ to $k=\kmax$.}
\tablenotetext{c}{$\ell=i$ stands for volume coverage at level $i$.  
The standard deviation of fluctuations in refinement coverage is $\la 3$\%}
\tablenotetext{d}{The total CPU time is normalized by the total CPU time of the Model 10 unigrid run.}
\tablenotetext{e}{Refinement ratio $n_{\rm ref}^0= 4$, corresponding to a maximum resolution of $512^3$.}
\tablenotetext{f}{Model 11 is a stress test of the code and has 
$\mrms=17.32$, $\beta_0=0.00667$ and $\mao=1$, in contrast to models 1-10, which have
$\mrms=10$, $\beta_0=0.1$ and $\mao=\sqrt{5}$.}
\end{deluxetable}
\clearpage


\begin{thebibliography}{}
\bibitem[Balsara \& Spicer(1999)]{bal99} Balsara, D. S. \& Spicer, D. S. 1999, J. Comput. Phys., 149, 270
\bibitem[Balsara(2001)]{bal01} Balsara, D. 2001, J. Comput. Phys., 174, 614
\bibitem[Batchelor(1953)]{bat53} Batchelor, G. K. 1953, Homogeneous Turbulence, Cambridge University Press.
\bibitem[Berger and Colella(1989)]{bc89} Berger M.J., Colella, P. 1989, J. Comput. Phys., 82, 64
\bibitem[Berthon(2005)]{ber05} Berthon, C. 2005, Commun. Math. Sci. 3 (2) 133
\bibitem[Brio \& Wu(1988)]{bri88} Brio, M. \& Wu, C. C. 1988, J. Comput. Phys., 75, 400
\bibitem[Colella(1990)]{col90} Colella, P. 1990, J. Comput. Phys., 87, 171
\bibitem[Colella et al.(2000)]{col00} Colella, P., Graves, D. T., Keen, N. D., et al. 2000, https://seesar.lbl.gov/ANAG/chombo)
\bibitem[Colella \& Woodward(1984)]{col84} Colella, P. \& Woodward, P. R. 1984, J. Comput. Phys., 54, 174
\bibitem[Collins et al.(2011)]{col11} Collins, D. C., Padoan, P., Norman, M. L., \& Xu, H. 2011, \apj, 731, 59
\bibitem[Crockett et al.(2005)]{cro05} Crockett, R. K., Colella, P., Fisher, R. T., Klein, R. I., \& McKee, C. F. 2005, J. Comput. Phys., 203, 422
\bibitem[Crutcher(1999)]{cru99} Crutcher, R. M. 1999, \apj, 520, 706
\bibitem[Dedner et al.(2002)]{ded02} Dedner, A., Kemm, F., Kr\"{o}ner, D., et al. 2002, J. Comput. Phys. 175, 645.
\bibitem[Evans \& Hawley(1988)]{eva88} Evans, C. R. \& Hawley, J. F. 1988, \apj, 659
\bibitem[Federrath et al.(2010)]{fed10} Federrath, C., Roman-Duval, J., Klessen, R. S., Schmidt, W., \& Mac Low, M.-M. 2010, \aap, 512, 81
\bibitem[Fromang et al.(2006)]{fro06} Fromang, S. Hennebelle, P., \& Teyssier, R. 2006, \aap, 457, 371
\bibitem[Fryxell et al.(2000)]{fry00} Fryxell, B., Olson, K., Ricker, P., et al. 2000, \apjs, 131, 273
\bibitem[Gardiner \& Stone(2005)]{gar05} Gardiner, T. A., \& Stone, J. M. 2005, J. Comp. Phys., 205, 509
\bibitem[Gardiner \& Stone(2008)]{gar08} Gardiner, T. A., \& Stone, J. M. 2008, J. Comp. Phys., 227, 4123
\bibitem[Gottlieb \& Shu(1996)]{got96} Gottlieb, S., Shu, C.-W. 1996, NASA CR-201591 ICASE Rep. 96-50, 20 (Washington: NASA)
\bibitem[Hayes et al.(2006)]{hay06} Hayes, J. C., Norman, M. L., Fiedler, R. A., et al. 2006, \apjs, 165, 188
\bibitem[Iroshnikov(1963)]{iro63} Iroshnikov, P. S. 1963, AZh, 40, 742 (English transl. Soviet Astron., 7, 566 [1964])
\bibitem[Kaneda et al.(2003)]{kan03} Kaneda Y, Ishihara T, Yokokawa M, Itakura K, \& Uno A. 2003, Phys. Fluids, 15, L21
\bibitem[Klein (1999)]{kle99} Klein, R. I. 1999, J. Comput. Appl. Math., 109, 123
\bibitem[Kraichnan(1965)]{kra65} Kraichnan, R. H. 1965, Phys. Fluids, 8, 1385
\bibitem[Kritsuk et al.(2006)]{kri06} Kritsuk, A. G., Norman, M. L., \& Padoan, P. 2006, \apjl, 638, L25
\bibitem[Kritsuk et al.(2007)]{kri07} Kritsuk, A. G., Norman, M. L., Padoan, P. \& Wagner, R. 2007, \apj, 665, 416
\bibitem[Kritsuk et al.(2009a)]{kri09a} Kritsuk, A. G., Ustyugov, S. D., Norman, M. L., \& Padoan, P. 2009a, J. Phys.: Conf. Ser., 180, 012020
\bibitem[Kritsuk et al.(2009b)]{kri09b} Kritsuk, A.~G., Ustyugov, S.~D., Norman, M.~L., \& Padoan, P.\ 2009b, Numerical Modeling of Space Plasma Flows: ASTRONUM-2008, 406, 15 
\bibitem[Krumholz et al.(2004)]{kru04} Krumholz, M. R., McKee, C. F., \& Klein, R. I. 2004, \apj, 611, 399
\bibitem[Krumholz et al.(2007)]{kru07} Krumholz, M. R., Klein, R. I., McKee, C. F., \& Bolstad, J. 2007, \apj, 667, 626
\bibitem[Lemaster \& Stone(2009)]{lem09} Lemaster, M. N. \& Stone, J. M. 2009, \apj, 691, 1092
\bibitem[LeVeque(1992)]{lev92} LeVeque, R. J., Numerical Methods for Conservation Laws, second ed., Birkhäuser, Basel, Switzerland, Boston, USA, 1992.
\bibitem[Li et al.(2008)]{li08} Li, P.S., McKee, C. F., \& Klein, R. I. 2008, \apj, 684, 380
\bibitem[Londrillo \& del Zanna(2004)]{lon04} Londrillo, P. \& del Zanna, L. 2004, J. Comput. Phys., 195, 17
\bibitem[Mac Low(1999)]{mac99} Mac Low, M.-M. 1999, 524, 169
\bibitem[Mac Low \& Klessen(2004)]{mac04} Mac Low, M.-M. \& Klessen, R. S. 2004, Rev. Mod. Phys., 76, 125
\bibitem[Martin \& Colella(2000)]{mar00} Martin, D \& Colella, P. 2000, J. Comput. Phys., 163, 271
\bibitem[Masunaga et al.(1998)]{mas98} Masunaga, H., Miyama, 
S.~M., \& Inutsuka, S.-I.\ 1998, \apj, 495, 346 
\bibitem[McKee \& Ostriker(2007)]{mck07} McKee, C. F., \& Ostriker, E. C. 2007, \araa, 45, 565
\bibitem[Mignone(2005)]{mig05} Mignone, A. 2005, J. Comput. Phys., 225, 1472
\bibitem[Mignone et al.(2007)]{mig07} Mignone, A., Bodo, G., Massaglia, S., et al. 2007, \apjs, 170, 228
\bibitem[Mignone \& Tzeferacos(2010)]{mig10} Mignone, A. \& Tzeferacos, P. 2010, J. of Comput. Phys., 229, 2117
\bibitem[Miyoshi \& Kusano(2005)]{miy05} Miyoshi, T., \& Kusano, K. 2005, J. of Comput. Phys., 208, 315
\bibitem[Peng \& Abel (2009)]{pen09} Wang, P. \& Abel, T. 2009, \apj, 696, 96
\bibitem[Powell et al.(1999)]{pow99} Powell, K. G., Roe, P. L., Linde, T. J., Gombosi, T. I., \& de Zeeuw, D. L. 1999, J. Comput. Phys., 153, 284
\bibitem[Roe(1986)]{roe86} Roe, P. L. 1986, Annu. Rev. Fluid Mech., 18, 337
\bibitem[Ryu \& Jones(1995)]{ryu95} Ryu, D. \& Jones, T. W. 1995, \apj, 442, 228
\bibitem[Schmidt et al.(2009)]{sch09} Schmidt, W., Federrath, C., Hupp, M., Kern S., \& Niemeyer, J. C. 2009, \aap, 494, 127
\bibitem[Stone et al.(2008)]{sto08} Stone, J. M., Gardiner, T. A., Teuben, P., Hawley, J. F., \& Simon, J. B. 2008, \apj, 178, 137
\bibitem[Stone et al.(1998)]{sto98} Stone, J. M., Ostriker, E. C., \& Gammie, C. F. 1998, \apjl, 508, L99
\bibitem[Toro(1999)]{tor99} Toro, E. F. 1999, Riemann Solvers and Numerical Methods for  Fluid Dynamics, (Berlin: Springer Verlag)
\bibitem[T\'{o}th(2000)]{tot00} T\'{o}th, G. 2000, J. Comput. Phys. 161 (2000) 605.
\bibitem[Truelove et al.(1997)]{tru97} Truelove, J. K., Klein, R. I., McKee, C. F., et al. 1997, \apj, 489, L179
\bibitem[Truelove et al.(1998)]{tru98} Truelove, J. K., Klein, R. I., McKee, C. F., et al. 1998, \apj, 495, 821
\bibitem[van Leer(1974)]{vanl74} van Leer, B. 1974, J. Comput. Phys., 14,
  361
\bibitem[Teyssier(2002)]{tey02} Teyssier, R. 2002, \aap, 385, 337
\bibitem[Verma(2007)]{ver07} Verma, M. K. \& Donzis, D. 2007, J. Physics A: Mathematical and Theoretical, 40, 4401
\bibitem[Waagan(2009)]{waa09} Waagan, K. 2009, J. of Comput. Phys., 228, 8609
\bibitem[Waagan et al.(2011)]{waa11} Waagan, K., Federrath, C. \& Klingenberg, C., 2011, J. of Comput. Phys., 230, 3331
\bibitem[Wang \& Abel(2009)]{wan09} Wang, P. \& Abel, T. 2009, \apj, 696, 96
\bibitem[Zhang \& Shu(2010)]{zha10} Zhang, X. \& Shu, C. W. 2010, J. Comput. Phys.,
229, 8918
\end{thebibliography}
\end{document}